\documentclass[twocolumn,english,reprint,superscriptaddress,longbibliography,pra]{revtex4-1}

\usepackage[T1]{fontenc}
\usepackage[latin9]{inputenc}
\usepackage{color}
\usepackage{bm}
\usepackage{amstext}
\usepackage{amssymb}
\usepackage{graphicx}
\usepackage{booktabs}
\usepackage{tabularx}
\usepackage{amsmath,mathrsfs}
\usepackage{enumitem}
\makeatletter
\usepackage{multibib}
   
\usepackage{babel}
\usepackage{bbold}
\usepackage{mathrsfs}
\usepackage{hyperref}
\hypersetup{
 linktocpage = true,
     colorlinks,
    citecolor=blue,
    filecolor=black,
    linkcolor=blue,
    urlcolor=blue
}

\newcommand{\id}{\mathbb{1}}
\usepackage{wasysym}
\usepackage{mathtools}

\newcommand{\tr}{\textrm{Tr}}

\newcommand{\ketbra}[2]{| #1 \rangle\!\langle #2 |}

\makeatother

\usepackage[normalem]{ulem}

\begin{document}
\author{Giacomo Torlai}
\email{gtorlai@flatironinstitute.org}
\affiliation{Center for Computational Quantum Physics, Flatiron Institute, New York, NY 10010, USA}

\author{Christopher J. Wood}
\affiliation{IBM T.J. Watson Research Center, Yorktown Heights, NY 10598, USA}

\author{Atithi Acharya}
\affiliation{Center for Computational Quantum Physics, Flatiron Institute, New York, NY 10010, USA}
\affiliation{Physics and Astronomy Department, Rutgers University, Piscataway, NJ 08854, USA}

\author{Giuseppe Carleo}
\affiliation{Center for Computational Quantum Physics, Flatiron Institute, New York, NY 10010, USA}

\author{Juan Carrasquilla}
\affiliation{Vector Institute, MaRS Centre, Toronto, Ontario, M5G 1M1, Canada}

\author{Leandro Aolita}
\affiliation{Instituto de F\'isica, Federal University of Rio de Janeiro, 21941-972, P. O. Box 68528, Rio de Janeiro, Brazil}

%%%%%%%%%%%%%%%%%%%%%%%%%%%%%%%%%%%%%%%%%%%%%%%%%%%%%%%%%%%%%%
\title{Quantum process tomography with unsupervised learning and tensor networks}

%%%%%%%%%%%%%%%%%%%%%%%%%%%%%%%%%%%%%%%%%%%%%%%%%%%%%%%%%%%%%%
\begin{abstract}
The impressive pace of advance of quantum technology calls for robust and scalable techniques for the characterization and validation of quantum hardware. Quantum process tomography, the reconstruction of an unknown quantum channel from measurement data, remains the quintessential primitive to completely characterize quantum devices. However, due to the exponential scaling of the required data and classical post-processing, its range of applicability is typically restricted to one- and two-qubit gates. Here, we present a new technique for performing quantum process tomography that addresses these issues by combining a tensor network representation of the channel with a data-driven optimization inspired by unsupervised machine learning. We demonstrate our technique through synthetically generated data for ideal one- and two-dimensional random quantum circuits of up to 10 qubits, and a noisy 5-qubit circuit, reaching process fidelities above 0.99 using only a limited set of single-qubit measurement samples and input states. Our results go far beyond state-of-the-art, providing a practical and timely tool for benchmarking quantum circuits in current and near-term quantum computers. 
\end{abstract}
%%%%%%%%%%%%%%%%%%%%%%%%%%%%%%%%%%%%%%%%%%%%%%%%%%%%%%%%%%%%%%

\maketitle

%%%%%%%%%%%%%%%%%%%%%%%%%%%%%%%%%%%%%%%%%%%%%%%%%%%%%%%%%%%%%%

{\bf Introduction.} Digital quantum computers and analog quantum simulators are entering regimes outside the reach of classical computing hardware~\cite{Preskill2018}. Coherent manipulation of complex quantum states with dozens of qubits have been realized across several platforms, including trapped ions~\cite{Smith:2016aa,Friis18}, Rydberg atom arrays~\cite{Bernien:2017aa}, cold atoms in optical lattices~\cite{Trotzky:2012aa}, and super-conducting qubit circuits~\cite{Arute:2019aa}. Over the next few years, it is expected that quantum devices will attain hundreds of qubits, unlocking a variety of quantum computing applications with far-reaching scientific and technological ramifications.

As the size and complexity of quantum hardware continues to grow, techniques capable of characterizing complex multi-qubit error processes are essential for developing error mitigation for near-term applications~\cite{Kandala:2017aa,Kokail:2019aa,Havlicek:2019aa,Arute2}. 
Recent efforts have focused on generalizations of randomized benchmarking~\cite{Magesan2012} to recover partial information about the strength and locality of correlated errors in larger devices~\cite{Erhard2019,Harper2019,Mckay2020}. However these approaches are restricted to non-universal gate-sets, and cannot be directly applied to generic quantum circuits. Other approaches exist for the validation of average fidelities of a prepared quantum state through a reduced set of measurements~\cite{Flammia11,daSilva11,Aolita15,Gluza18,Roth18}, but only provide limited information about the nature of the noise in the preparation circuit.

The gold standard for the full characterization of quantum gates and circuits is quantum process tomography (QPT)~\cite{Chuang97,DAriano:2001aa}, a procedure that reconstructs an unknown quantum process from measurement data. A direct approach to QPT relies on a informationally-complete (IC) set of measurement settings, which inevitably leads to an algorithmic complexity -- in terms of number of measurements and classical post-processing -- that scales exponentially with the number of qubits. Due to these limitations, QPT has only been experimentally implemented on up to 3 qubits~\cite{Obrien04,Riebe2006,Weinstein2004,Bialczak2010,Chow2012,Shabani11,Krinner20,Govia20}.

In most practical scenarios, however, a process to be characterized in a quantum computer typically contains structure, which can stem from the restricted set of operations in an experiment or the details and severity of the inherent noise in the device. This suggests that it may be possible to accurately describe certain relevant quantum channels by means of classical resources with only polynomial overhead. This insight has been leveraged successfully in quantum state tomography, the data-driven reconstruction of a quantum state. Notable examples include matrix product state (MPS) tomography~\cite{cramer2009efficient,MPOtomo,Lanyon2017}, exploiting low-entanglement representations of quantum states, and compressed sensing~\cite{Gross:2010prl,Shabani11}, relying on the assumption of sparsity of the  measurement data. 
 
More recently, an alternative theoretical framework for quantum state tomography based on machine learning has been put forward ~\cite{Torlai:2020aa,torlai_2018_nnqst,Carrasquilla:2019aa}, and implemented in a cold-atom experiment~\cite{Torlai19}. This approach leverages the effectiveness of unsupervised machine learning in extracting high-dimensional probability distributions from raw data~\cite{Goodfellow-et-al-2016}, combined with the high expressivity of neural networks for capturing highly-entangled quantum many-body states~\cite{Carleo602,2017NatCoGAO,PhysRevX.8.011006,Carleo:2018aa}. In contrast, the development of approximate algorithms for QPT applicable to near-term quantum devices is currently lacking. While progress has been made in the context for learning non-Markovian dynamics~\cite{guo2020tensor,white2020experimental}, the question of a scalable method capable of reconstructing noisy quantum circuits remains wide open.

In this work, we present a novel technique to perform QPT of quantum circuits of sizes well beyond state-of-the-art.  By exploiting the structure of the problem, our approach alleviates important scaling issues of standard QPT. We combine elements of two state-of-the-art classes of algorithms, namely a tensor-network representation of a quantum channel and a data-driven global optimization inspired by unsupervised learning algorithms. We show numerical experiments on synthetic data for unitary circuits, reaching reconstruction fidelities above 0.99 for a 10-qubit depth-4 random quantum circuit using less than $10^5$ single-shot measurements out of the tomographycally complete set of $\sim\!\!10^{12}$ settings. We also demonstrate the reconstruction of a single 5-qubit parity-check measurement in the surface code undergoing an amplitude damping noise channel. Our proposed method paves the way to the robust and scalable verification of quantum circuits implemented in current experimental hardware.

%%%%%%%%%%%%%%%%%%%%%%%%%%%%%%%%%%%%%%%%%%%%%%%%%%%%%%%%%%%%%%

{\bf Quantum process tomography.} 
A general $N$-qubit quantum channel is described by a completely-positive (CP) trace-preserving (TP) map $\mathcal{E}$. There exist several equivalent mathematical representations of a quantum channel~\cite{Wood:2015qic}, and in the context of process tomography, it is most natural to use the {\it Choi matrix} representation~\cite{Jamiokowski:1972aa, Choi1975}. The Choi matrix is a positive semidefinite operator
\begin{equation}
\bm{\Lambda}_{\mathcal{E}} = \big(\mathbb{1}\otimes\mathcal{E}\big) \big(|\Phi^+\rangle\langle\Phi^+|^{\otimes N}\big)\:,
\label{Eq::ChoiDef}
\end{equation}
where $\mathcal{E}$ is applied to one half of the tensor product of $N$ unnormalized Bell pairs $|\Phi^+\rangle=|00\rangle+|11\rangle$. The channel $\mathcal{E}$ is CP if and only if the Choi-matrix is positive-semidefinite ($\bm{\Lambda}_{\mathcal{E}} \ge 0$), and TP if and only if the partial trace over the subspace of $N$ qubit acted upon by the channel $\mathcal{E}$ in Eq.~(\ref{Eq::ChoiDef}) is the identity matrix~\cite{Wood:2015qic}. It follows that $\bm{\Lambda}_{\mathcal{E}}$ is isomorphic to an unnormalized density operator over an extended (bipartite) $2N$-qubit Hilbert space ($\text{Tr}\,\bm{\Lambda}_{\mathcal{E}}=d^N$, with $d$ the dimension of the local Hilbert space, i.e. $d=2$ for qubits).

Because of the one-to-one correspondence between the Choi matrix and the map $\mathcal{E}$, QPT reduces to the data-driven reconstruction of $\bm{\Lambda}_{\mathcal{E}}$. The standard approach to QPT consists of fitting the matrix elements of $\bm{\Lambda}_{\mathcal{E}}$ (parametrized in full), typically using convex optimization techniques, from the statistics of an IC set of measurements on the output state, applied to an IC set of input quantum states $\{\bm{\rho}_i\}$. The major limitation of full QPT is that the size of the preparation and measurement sets scales exponentially with the number of qubits.

Our approach to overcome the limitations of full QPT relies on two ingredients. First, an efficient representation of a Choi matrix in terms of a tensor network, whose total number of parameters is small compared to the dimension of the process Hilbert space. Second, an unsupervised learning algorithm to discover an optimal set of tensor-network parameters, which consists of minimizing the statistical divergence between the corresponding process probability distribution and the one underlying the measurement data. Most importantly, if the unknown quantum channel possesses enough structure, and the tensor-network parametrization has enough representational power, unsupervised learning can allow the model to generalize beyond the acquired measurements, enabling the process reconstruction using a reduced number of the state preparation and measurement bases required for standard QPT. The unsupervised learning optimization, in stark contrast with reconstructions using linear inversion~\cite{Chuang97,DAriano:2001aa} or MPS tomography~\cite{cramer2009efficient}, is at the heart of the scalability of our method.

\begin{figure}[t!]
\noindent \centering \includegraphics[width=\columnwidth]{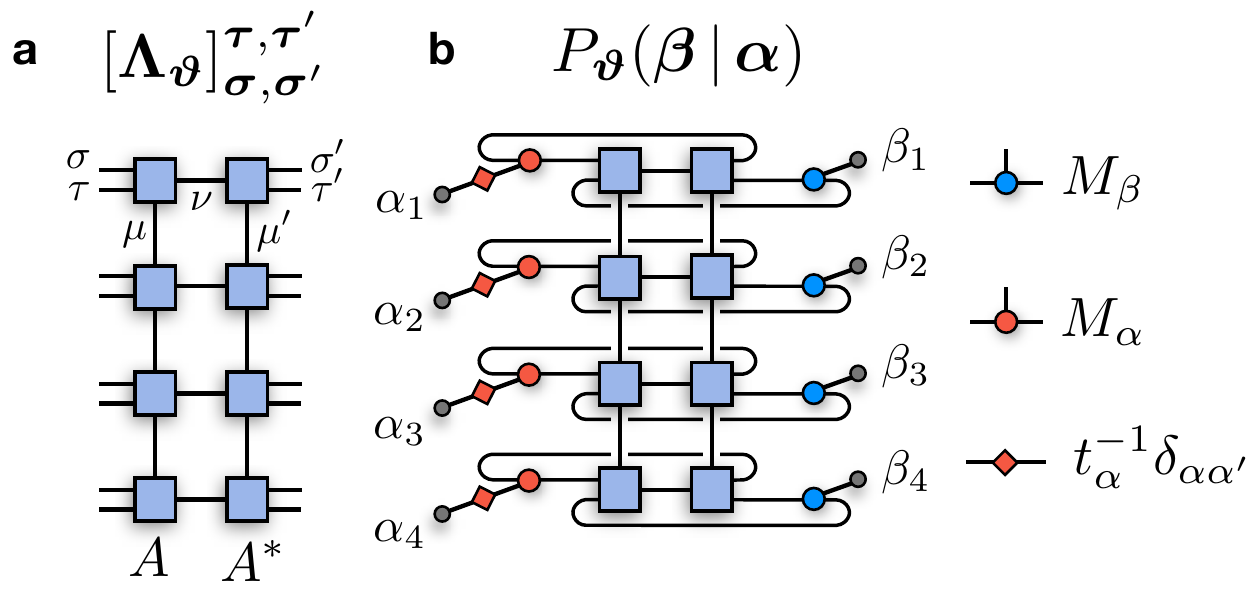}
\caption{{\bf Quantum process tomography with tensor networks}. ({\bf a}) The quantum process is represented by a Choi matrix $\bm{\Lambda}_{\bm{\vartheta}}$, parametrized by a locally-purified density operator (LPDO). The input and output indices of the process are $\{\sigma_j\}$ and $\{\tau_j\}$ respectively. ({\bf b}) Tensor contraction evaluating the conditional probability distribution $P_{\bm{\vartheta}}(\bm{\beta}\,|\,\bm{\alpha})$, i.e. the probability that the LPDO Choi matrix associates with the measurement $\bm{M}_{\bm{\beta}}$ given the state $\bm{\rho}_{\bm{\alpha}}=t_{\bm{\alpha}}^{-1}\bm{M}_{\bm{\alpha}}$ at the input of the channel.}
\label{Fig::1}
\end{figure}

{\bf Tensor-network unsupervised learning.}  
We begin by introducing a parametrization of the Choi matrix $\bm{\Lambda}_{\bm{\vartheta}}$ (with $\bm{\vartheta}$ the set of variational parameters) in terms of {\it locally-purified density operator} (LPDO), a class of matrix product operators that are non-negative by construction~\cite{Werner:2016aa} (Fig.~\ref{Fig::1}a). Given a basis for the input $\{|\bm{\sigma}\rangle\}$ and the output $\{|\bm{\tau}\rangle\}$ Hilbert spaces of the channel, the matrix elements $\langle\bm{\sigma},\bm{\tau}|\bm{\Lambda}_{\bm{\vartheta}}|\bm{\sigma}^\prime,\bm{\tau}^\prime\rangle$ of the LPDO Choi matrix are given by
\begin{equation}
[\bm{\Lambda}_{\bm{\vartheta}}]^{\bm{\tau},\bm{\tau}^\prime}_{\bm{\sigma},\bm{\sigma}^\prime}=\sum_{\{\bm{\mu},\bm{\mu}^\prime\}}\sum_{\{\bm{\nu}\}}\:
\prod_{j=1}^N\:[A_j]^{\tau_j,\sigma_j}_{\mu_{j-1},\nu_j,\mu_{j}}[A^*_j]^{\tau^\prime_j,\sigma^\prime_j}_{\mu^\prime_{j-1},\nu_j,\mu^\prime_{j}}\:,
\label{Eq::LPDO}
\end{equation}
where $\bm{\vartheta}=\{A_j\}$. Here, we assume that $\{A_j\}$ already incorporate the proper normalization $\text{Tr}_{\bm{\sigma},\bm{\tau}}\:\bm{\Lambda}_{\bm{\vartheta}}=d^N$. Each tensor $A_j$ has input index $\sigma_j$, output index $\tau_j$, {\it bond indices} $(\mu_{j-1},\mu_{j})$ and {\it Kraus index} $\nu_j$. The bond and Kraus dimensions of the LPDO are defined as $\chi_\mu=\max_j\{\chi_{\mu_j}=\text{dim}[\mu_j]\}$ and $\chi_\nu=\max_j\{\chi_{\nu_j}=\text{dim}[\nu_j]\}$. By setting $\chi_\nu=1$, the resulting rank-1 Choi matrix is $\bm{\Lambda}_{\bm{\vartheta}}=|\bm{\Psi}_{\bm{\vartheta}}\rangle\!\langle\bm{\Psi}_{\bm{\vartheta}}|$, where $|\bm{\Psi}_{\bm{\vartheta}}\rangle$ is an MPS with physical dimension $d^2$ and bond dimension $\chi_\mu$.

To perform process tomography with LPDOs, we consider the standard QPT setup of positive operator valued measures (POVM) $\bm{M}_{\bm{\beta}}=\bigotimes_{j=1}^NM_{\beta_j}$, where $\{M_{\beta_j}\}_{\beta_j=1}^{K_m}$ are single-qubit POVMs with $K_m$ measurement outcomes (${M}_{\beta_j}\ge0$ and $\sum_{\beta_j} M_{\beta_j}=\mathbb{1}_j$). As input states to the channel, we take product states $\bm{{\rho}}_{\bm{\alpha}}=\bigotimes_{j=1}^N\rho_{\alpha_j}$ with $\alpha_j=1,\dots,K_p$. The preparation states and output measurements are  identified by the classical strings $\bm{\alpha}=(\alpha_1,\dots,\alpha_N)$ and $\bm{\beta}=(\beta_1,\dots,\beta_N)$ respectively. The output state of the channel is obtained from the Choi matrix as~\cite{Wood:2015qic}
\begin{equation}
\mathcal{E}(\bm{\rho}_{\bm{\alpha}})=\text{Tr}_{\bm{\sigma}}\,\Big[(\bm{\rho}_{\bm{\alpha}}^{T}\otimes\mathbb{1}_{\bm{\tau}})\bm{\Lambda}_{\mathcal{E}}\Big]\:,
\end{equation}
where $\bm{\rho}_{\bm{\alpha}}^{T}$ stands for matrix transposition. For any state $\mathcal{E}(\bm{\rho}_{\bm{\alpha}})$, the probability that a POVM measurement yields outcome $\bm{M}_{\bm{\beta}}$ is
\begin{equation}
P_{\mathcal{E}}(\bm{\beta}\,|\,\bm{\alpha})=\text{Tr}_{\bm{\sigma},\bm{\tau}}\,\Big[(\bm{\rho}_{\bm{\alpha}}^{T}\otimes\bm{M}_{\bm{\beta}})\bm{\Lambda}_{\mathcal{E}}\Big]\:.
\end{equation}
As long as the input states and output POVM set are IC, the conditional probability distribution $P_{\mathcal{E}}(\bm{\beta}\,|\,\bm{\alpha})$ uniquely characterizes the channel $\mathcal{E}$, and provides a direct link between measurement statistics and the Choi matrix.

In the following, we use for convenience an over-complete set of input states $\bm{{\rho}}_{\bm{\alpha}}=t_{\bm{\alpha}}^{-1}\bm{{M}}_{\bm{\alpha}}$ (i.e. $K_m=K_p\equiv K$), where $ t_{\bm{\alpha}}=\text{Tr}\,\bm{{M}}_{\bm{\alpha}}=\prod_j\text{Tr}\,M_{\alpha_j}$ is a normalization factor. To generate a training data set, we prepare a finite set of $M$ input states $\{\bm{\rho}^{(k)}_{\bm{\alpha}}\}_{k=1}^{M}$, randomly sampled according to a fixed prior distribution $Q(\bm{\alpha})$. We then apply the channel to each state, and perform a measurement at its output, recording the outcomes $\{\bm{M}^{(k)}_{\bm{\beta}}\}_{k=1}^{M}$. The resulting data set is specified by $M$ strings of $2N$ $K$-valued integers, $\mathcal{D}=\{(\bm{\alpha}^{(k)},\bm{\beta}^{(k)})\}_{k=1}^{M}$, with  joint probability distribution $\break P_{\mathcal{D}}(\bm{\alpha},\bm{\beta})=Q(\bm{\alpha})P_{\mathcal{E}}(\bm{\beta}\,|\,\bm{\alpha})$. Similarly, we can estimate the corresponding probability distribution $P_{\bm{\vartheta}}(\bm{\beta}\,|\,\bm{\alpha})$ for the Choi matrix $\bm{\Lambda}_{\bm{\vartheta}}$. Since both input states and output POVMs factorize over the extended Hilbert space, estimating the probability translates into local contractions of the tensors $A_j$ with the tensor product $\rho^T_{\alpha_j}\otimes M_{\beta_j}$ at all sites $j$ (Fig.~\ref{Fig::1}b). The cost of this operation is $\mathcal{O}(d^2N\chi_\nu\chi_\mu^3)$, remaining efficient as long as the bond dimensions $(\chi_\mu,\chi_\nu)$ are sufficiently small.

The learning procedure, inspired by generative modeling of neural networks in machine learning applications~\cite{Goodfellow-et-al-2016}, consists of varying the parameters $\bm{\vartheta}$ to minimize the distance between the LPDO distribution $P_{\bm{\vartheta}}(\bm{\beta}\,|\,\bm{\alpha})$ and the target distribution $P_{\mathcal{E}}(\bm{\beta}\,|\,\bm{\alpha})$, averaged over the input prior $Q(\bm{\alpha})$. As a measure of probability distance, we adopt the Kullbach-Leibler divergence~\cite{Kullback:1951aa}:
\begin{equation}
D_{KL}=\sum_{\{\bm{\alpha}\}}Q(\bm{\alpha})\sum_{\{\bm{\beta}\}} P_{\mathcal{E}}(\bm{\beta}\,|\,\bm{\alpha})\log\frac{P_{\mathcal{E}}(\bm{\beta}\,|\,\bm{\alpha})}{P_{\bm{\vartheta}}(\bm{\beta}\,|\,\bm{\alpha})}\:,
\label{Eq::KL_full}
\end{equation}
Minimizing this quantity is equivalent to minimizing the negative-log likelihood
\begin{equation}
\mathcal{C}({\bm{\vartheta}})=-\frac{1}{M}\sum_{k=1}^{M}\log P_{\bm{\vartheta}}(\bm{\beta}_k\,|\,\bm{\alpha}_k)\:,
\label{Eq::NLL}
\end{equation}
where the average is taken over the data set $\mathcal{D}$. This is the cost function of our optimization problem. This type of tensor network optimization, also explored for quantum state tomography~\cite{PhysRevA.101.032321}, is in contrast with the local optimization used in the original formulation of MPS tomography, which relies on measurements of local subsystems and entails and exponential scaling with the size of the subsystems~\cite{Lanyon2017,Govia20}.

The LPDO parameters are iteratively updated using gradient descent $\bm{\vartheta}\rightarrow\bm{\vartheta}-\eta\:\nabla_{\bm{\vartheta}}\mathcal{C}({\bm{\vartheta}})$ (or a variation thereof), where $\eta$ is the size of the gradient update (i.e. the {\it learning rate}). In our simulations, we optimize the LPDO 
using automatic differentiation software~\cite{tensorflow}, a framework that is being increasingly explored in tensor networks applications~\cite{Liao:2019aa,torlai2019wavefunction}. However, the gradients of the cost function can also be derived analytically~\cite{Han:2018aa,glasser2019expressive}, and are shown in the Supplementary Material. 

In defining our parametrized model $\bm{\Lambda}_{\bm{\vartheta}}$, we exploited the fact that Choi matrices are isomorphic to density operators, which justifies the use of LPDOs. However, while $\bm{\Lambda}_{\bm{\vartheta}}=\bm{\Lambda}^\dagger_{\bm{\vartheta}}$ and $\bm{\Lambda}_{\bm{\vartheta}}\ge0$ by construction, the LPDO is inherently not TP. That is, the condition $\text{Tr}_{\bm{\tau}}\,\bm{\Lambda}_{\bm{\vartheta}}=\id_{\bm{\sigma}}$ is not enforced at the level of the elementary tensors $\{A_j\}$. We expect that, if $M$ is large enough and the model faithfully learns the quantum channel underlying the training data set, this property should also be satisfied. Nonetheless, we can approximately impose the TP constraint by adding a {\it regularization} term to $\mathcal{C}(\bm{\vartheta})$, which induces a bias towards trace-preserving matrices. We define this regularization term as 
\begin{equation}
\Gamma_{\bm{\vartheta}}= \sqrt{d^{-N}}\|\bm{\Delta}_{\bm{\vartheta}}\|_F=\sqrt{d^{-N}}\sqrt{\text{Tr}_{\bm{\sigma}}\big(\bm{\Delta}_{\bm{\vartheta}}\bm{\Delta}_{\bm{\vartheta}}^\dagger\big)}\:,
\end{equation}
where $\bm{\Delta}_{\bm{\vartheta}}=\text{Tr}_{\bm{\tau}}\bm{\Lambda}_{\bm{\vartheta}}-\id_{\bm{\sigma}}$. The final cost function becomes $\mathcal{C}({\bm{\vartheta}})=-\langle \log P_{\bm{\vartheta}}(\bm{\beta}\,|\,\bm{\alpha})\rangle_{\mathcal{D}}+\kappa\, \Gamma_{\bm{\vartheta}}$, where $\kappa$ is a hyper-parameter of the optimization.

{\bf Numerical experiments.} We study the performance of LPDO-based QPT for unitary and noisy quantum channels. We adopt, for both the input states and measurements, the IC-POVM set built out of the rank-1 projectors of the $K=6$ eigenstates of the Pauli matrices. For all the instances described, we generate the training data set $\mathcal{D}$ using a uniform prior distribution $Q(\bm{\alpha})=K^{-N}$. We split the data set into a training set and a validation set, containing respectively 80\% and 20\% of the total data. The training data set contains the measurements used to compute the gradients and train the LPDO. The remaining held-out data is used for cross-validation for selecting the optimal model, i.e. the set of parameters yielding the lowest value of the cost function computed on the validation data set. Details on the data generation and the LPDO trainings are provided in the Supplementary Material.

We start by studying the case of a unitary channel characterized by a rank-1 Choi matrix $\bm{\Lambda}_{\mathcal{E}}=|\bm{\Psi}_{\mathcal{E}}\rangle\!\langle\bm{\Psi}_{\mathcal{E}}|$. We perform QPT by setting the Kraus dimension to $\chi_\nu=1$, leading to the parametrized Choi matrix $\break\bm{\Lambda}_{\bm{\vartheta}}=|\bm{\Psi}_{\bm{\vartheta}}\rangle\!\langle\bm{\Psi}_{\bm{\vartheta}}|$ expressed in terms of an MPS $\bm{\Psi}_{\bm{\vartheta}}$. We also set the bond dimension of the LPDO $\chi_\mu$ equal to the bond dimension $\chi_{\mathcal{E}}$ of $\bm{\Psi}_{\mathcal{E}}$. Thus, there is no approximation in the representation of the channel, and any reconstruction error generates solely from the finite size of the data set and any potential inefficiency of the optimization procedure. We point out that, when the ideal target quantum circuit is known, it is possible to estimate what is the minimum value of $\chi_{\mathcal{E}}$ leading to a faithful tensor-network representation of the quantum circuit. Both conditions on $\chi_\mu$ and $\chi_\nu$ will be lifted for the reconstruction of a noisy channel, later in this section.

During the training, we measure the cost function computed on both the training and validation data sets. The former monitors the learning progress, while the latter monitors the overfitting and is used to select the optimal parameters. In addition, we also measure the {\it reconstruction fidelity}, which we defined as the {\it quantum process fidelity} $\mathcal{F}(\bm{\Lambda}_{\mathcal{E}},\bm{\Lambda}_{\bm{\vartheta}})$ of the reconstruction to the true channel used to generate the data. The process fidelity is equivalent to the quantum state fidelity between the two (properly normalized) Choi matrices
\begin{equation}
\mathcal{F}(\bm{\Lambda}_{\mathcal{E}},\bm{\Lambda}_{\bm{\vartheta}})=d^{-2N}
\bigg(\text{Tr}\:\sqrt{\sqrt{\bm{\Lambda}_{\mathcal{E}}}\bm{\Lambda}_{\bm{\vartheta}}\sqrt{\bm{\Lambda}_{\mathcal{E}}}}\bigg)^2\:.
\end{equation} 
Note that, while this measurement cannot be performed in a scalable manner for two arbitrary (noisy) Choi-matrices, it is also not useful in an experimental scenario where $\bm{\Lambda}_{\mathcal{E}}$ is unknown. In this situation, one typically compares the fidelity between the reconstructed and the \emph{expected} ideal unitary channel, $\break\mathcal{F}(\bm{\Lambda}_{\bm{\vartheta}},\bm{\Lambda}_{\mathcal{E}})=d^{-2N}
\langle\bm{\Psi}_{\mathcal{E}}|\bm{\Lambda}_{\bm{\vartheta}}|\bm{\Psi}_{\mathcal{E}}\rangle$, which can be carried out using LPDOs by a tensor contraction. Here we consider the former definition, as we are benchmarking the faithfulness of the reconstruction.
\begin{figure}[t!]
\noindent \centering \includegraphics[width=1\columnwidth]{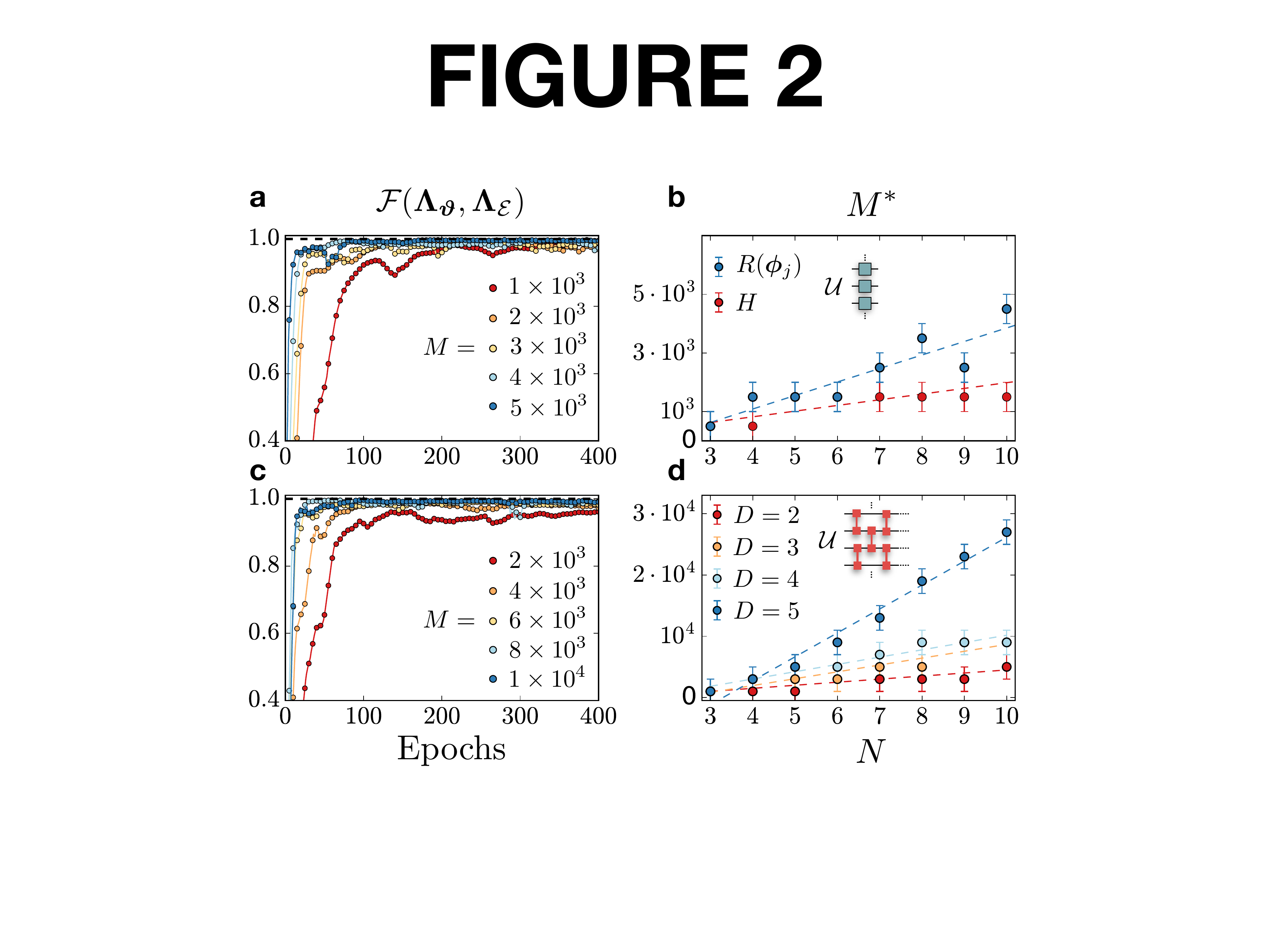}
\caption{{\bf Benchmarking unitary circuits.} We show the process reconstruction for unitary quantum circuits containing single-qubit and two-qubit quantum gates. ({\bf a}) Reconstruction fidelity during the LPDO training for a circuits with $N=4$ qubits containing a single layer of Hadamard gates. Different curves corresponds to an increasing size $M$ of the data set. ({\bf b}) Scaling of the minimum number of samples $M^*$ as a function of $N$ to reach a reconstruction infidelity of $\varepsilon=0.025$ (i.e. the sample complexity) for a circuit with Hadamard gates (red) and a circuit with random single-qubit rotations $R(\bm{\varphi}_j)$ (blue). ({\bf c}) Reconstruction fidelity for a circuit with $N=4$ qubits containing 4 layers of controlled-not (CX) gates, for various data set sizes $M$. ({\bf d}) Sample complexity for quantum circuits with different depths $D$ containing layers of CX gates. For the sample complexity plots, the value $M^*$ is obtained by sequentially increasing $M$ until the threshold in accuracy is met. Error bars are given by the step-size in $M$, and dashed lines are linear fits.}
\label{Fig::2}
\end{figure}

\begin{figure*}[t]
\noindent \centering \includegraphics[width=2\columnwidth]{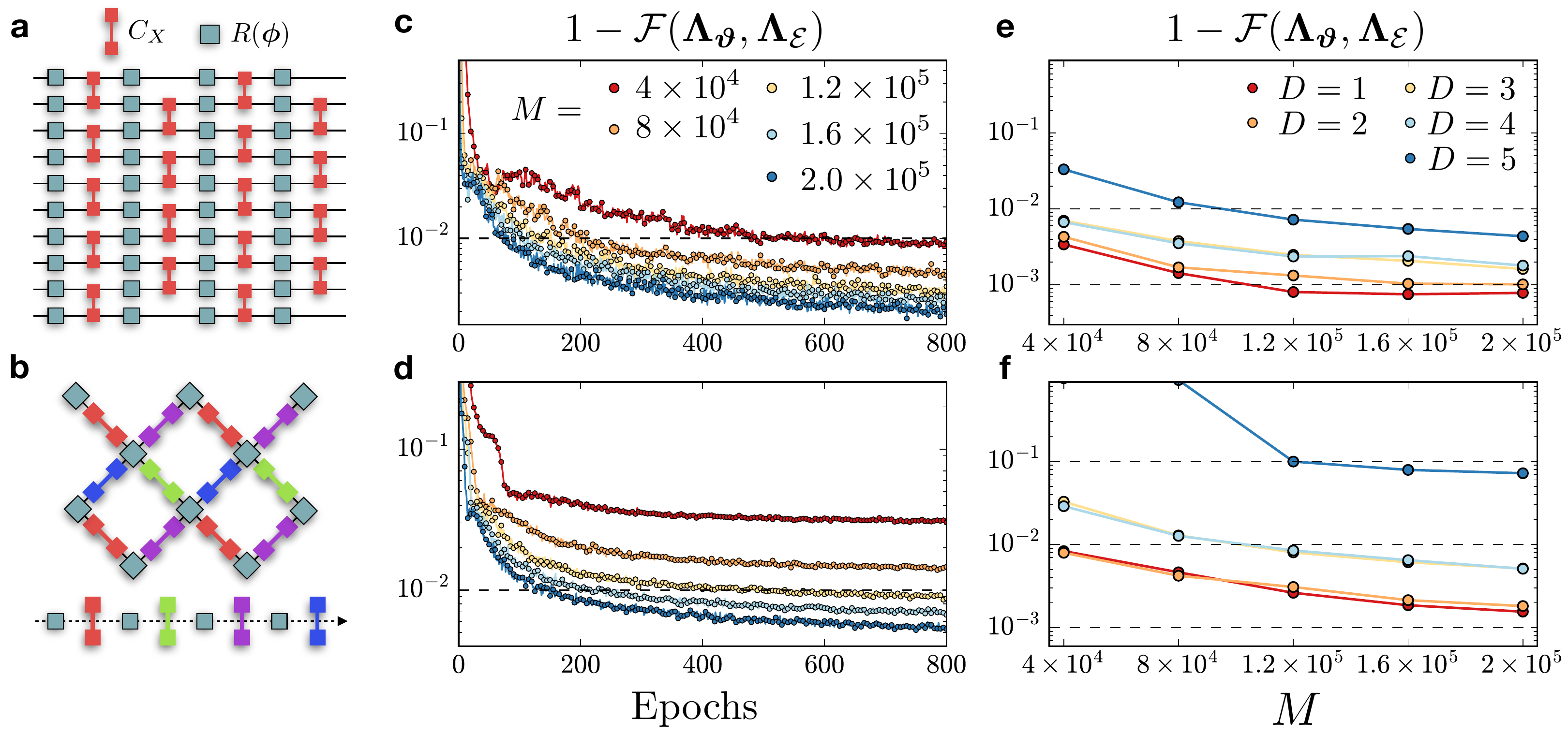}
\caption{{\bf Random quantum circuits}. ({\bf a}) One-dimensional quantum circuit with $N=10$ qubits and $D=4$ layers, each one consisting of random single-qubit rotations and CX gates, the latter applied in a staggered pattern between even and odd layers. ({\bf b}) Two-dimensional random quantum circuit, where each layer applies random single-qubit rotations and CX gates according to the colored sequence shown at the bottom of the image. In the subplots ({\bf c}) and ({\bf d}) we show the reconstruction infidelity at each epoch respectively for a one- and two-dimensional quantum circuit with depth $D=4$, for various data set sizes $M$. Subplots ({\bf e}) and ({\bf f}) show the lowest infidelities, obtained via cross-validation on held-out data, as a function of the data set size $M$ for different depths.}
\label{Fig::3}
\end{figure*}

The first test-case is a unitary quantum circuit containing a single layer of Hadamard gates acting on all qubits. We train LPDOs for different sizes $M$ of the training data set, and we show in Fig.~\ref{Fig::2}a the corresponding reconstruction fidelities measured at each training iteration ({\it epoch}), for $N=4$ qubits. Note that since the channel is noiseless, the reconstruction fidelity reduces to the  quantum state fidelity for pure states, $\mathcal{F}(\bm{\Lambda}_{\mathcal{E}},\bm{\Lambda}_{\bm{\vartheta}})=d^{-2N}|\langle\bm{\Psi}_{\mathcal{E}}|\bm{\Psi}_{\bm{\vartheta}}\rangle|^2$. From this data, we can compute the minimum number of training samples $M^*$ required to reach a fixed accuracy $\varepsilon$ in the reconstruction infidelity $1-\mathcal{F}(\bm{\Lambda}_{\bm{\vartheta}},\bm{\Lambda}_{\mathcal{E}})$. By repeating the same experiment for several systems sizes up to $N=10$ (with $\varepsilon=0.025$), we show the sample complexity  in Fig.~\ref{Fig::2}b --  the value $M^*$ as a function of $N$ -- observing a favorable scaling consistent with a linear behavior.  We repeat the same experiment for a single layer of random single-qubit rotations $R(\bm{\varphi}_j)$, observing a similar scaling with a steeper slope.

We also consider a quantum circuit containing $D$ layers of  controlled-NOT (CX) gates applied between neighboring qubits in a one-dimensional geometry. Each layer is applied in a staggered manner (inset of Fig.~\ref{Fig::2}d). We perform the same analysis as for the one-qubit circuits, and plot the fidelity curves for various $M$ for a circuit with $N=4$ qubits and depth $D=4$ (Fig.~\ref{Fig::2}c). The sample complexity, computed in an analogous manner, is shown in Fig.~\ref{Fig::2}d for different depths $D$. As expected, the threshold $M^*$ increases with the depth of the circuit. 

We now move to a more challenging case, and reconstruct a 10-qubit random quantum circuit with depth $D$ for both one- and two-dimensional qubit arrays. Each layer in the circuit consists of $N$ random single-qubit rotations followed by a layer of CX gates. For the one-dimensional circuit, the CX gates alternates between even and odd layers (Fig.~\ref{Fig::3}a). For the two-dimensional circuit, the CX gates are applied in a sequence according to the colors shows in Fig.~\ref{Fig::3}b. In the plots of Fig.~\ref{Fig::3}c-d we show the process infidelity during the training for depth-4 circuits and different values of the data set size $M$. We observe that, with enough number of single-shot samples $M$, the reconstructions surpass a fidelity of $\mathcal{F}=0.99$.

We evaluate the optimal LPDO parameters using cross-validation on the held-out data, a metric that does not rely on any prior information about the process and available in an experimental setting. We show in Fig.~\ref{Fig::3}e-f the corresponding lowest infidelities obtained during the training as a function of $M$. As in the previous case, the number of samples to reach a given accuracy increases with the depth of the circuit. For the one-dimensional circuit, the fidelity reach $\mathcal{F}>0.99$ with $4\times10^4$ measurements up to $D=4$, and converges to $\mathcal{F}\approx0.999$ and $\mathcal{F}\approx0.998$ for $D=2$ and $D=4$ respectively. For the two-dimensional circuit, the fidelity converges to $\mathcal{F}>0.99$ up to $D=4$ at $M=2\times10^5$, while $\mathcal{F}\approx0.93$ for $D=5$. In this case, the bond dimension of the target circuit is $\chi_\mu=32$, a four-fold increase from $\chi_\mu=8$ of the $D=4$ circuit. We emphasize that the data set size $M$ used is a very small fraction of the total number of input states and measurement settings. For a 10-qubit circuit, using the overcomplete set of preparation states, the total number of settings is $6^N3^N\sim10^{12}$.

Finally, we turn to the case of a quantum circuit undergoing a noise channel. As a test case, we study a single $X$-stabilizer measurement of the surface code, a paradigmatic model of topological quantum computation~\cite{Dennis:2002aa,Fowler:2012aa}. The circuit contains a total of $N=5$ qubits, where a single measurement qubit is used to stabilize the $X$ parity-check between four data qubits. The quantum circuit for the stabilizer measurement consists of a Hadamard gate on the measurement qubit, four CX gates between the measurement qubit and each data qubit, followed by an additional Hadamard gate on the measurement qubit. We apply a single-qubit amplitude damping channel to all gates with a fixed decay probability $\gamma\in[0,\dots,0.05]$.

\begin{figure}[t!]
\noindent \centering \includegraphics[width=1\columnwidth]{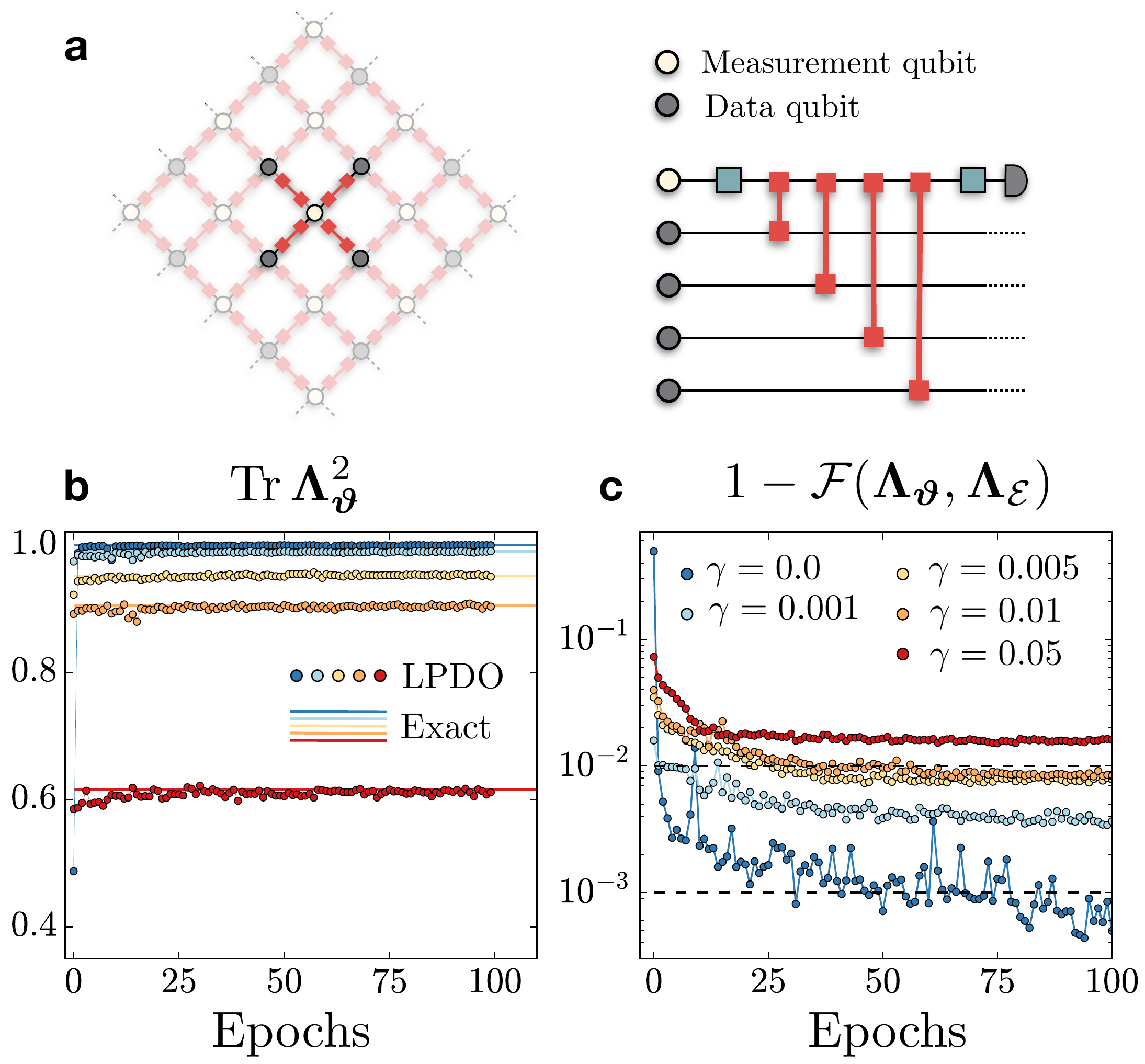}
\caption{{\bf Noisy stabilizer in the surface code}. ({\bf a}) A $X$-stabilizer plaquette embedded into the surface code (left) and the quantum circuit performing the parity-check measurement (right), containing Hadamard and CX gates. ({\bf b}) Purity of the LPDO Choi matrix during training (markers), compared to the exact Choi matrix (solid lines). ({\bf c}) Infidelity measurement during training for a data set size of $M=5\times10^5$ single-shot measurement outcomes.}
\label{Fig::4}
\end{figure}

We perform the reconstruction by varying both the bond dimension and the Kraus dimension, until convergence is found, and we show the results for $\chi_{\mu}=\chi_\nu=6$. During the training, we measure the reconstruction fidelity, as well as the purity $\tr\,\bm{\Lambda}^2_{\bm{\vartheta}}$ of the LPDO. For all values of the decay probability $\gamma$, we observe that the purity converges to the correct value (solid lines) computed from the exact Choi matrix (Fig.~\ref{Fig::4}b), suggesting that the Kraus dimension of the LPDOs is sufficient to capture the target noisy channel. We also show the process infidelity curves obtained using a total of $M=5\times10^5$ measurement samples, for different values of $\gamma$  (Fig.~\ref{Fig::4}c). While for the noiseless channel the fidelity reaches $\mathcal{F}>0.999$, the learning appears to become increasingly harder for larger values of $\gamma$. The lowest fidelity $\mathcal{F}\approx0.985$ is found at $\gamma=0.05$, which is a fairly large decay probability for current experiments. The reconstruction reaches $\mathcal{F}>0.99$ for the lower levels of noise.

{\bf Conclusions.} We introduced a procedure for quantum process tomography that integrates a tensor network representation of the Choi matrix in terms of a locally-purified matrix product operator~\cite{Werner:2016aa}, and an optimization strategy motivated by machine learning algorithms for generative modeling of high-dimensional probability distributions~\cite{Goodfellow-et-al-2016}. We demonstrated the power and scalability of the technique using simulated data for unitary random quantum circuits, reaching system sizes of up to 10 qubits and depth 5, and a stabilizer measurement of the surface code undergoing amplitude damping noise. In both cases, the resulting process fidelities reach values close to $\mathcal{F}=0.99$, using single-shot samples corresponding to a small fraction of the total number of preparation and measurements in the corresponding informationally-complete set, amenable to current experiments.

Due to the entanglement structure induced by the Choi matrix representation in terms of a tensor network with small bond dimension, our technique lends itself extremely well to the characterization of quantum hardware operating circuits of sufficiently low depth. The stringent limitation of standard process tomography in the accessible number of qubits are lifted, allowing the reconstruction of large quantum circuits for the case of one- (and quasi-one-) dimensional geometries. 

Our work demonstrates how infusing state-of-the-art tensor network algorithms with  machine learning ideas has the potential to unlock progress in the validation and characterization of currently available quantum devices, and in the design of better error mitigation protocols. This combination elevates quantum process tomography to a scale relevant for the solution of several key obstacles to realizing large-scale quantum computation such as the need for quantum error correction and fault tolerance, which naturally calls for the systematic benchmarking of large quantum circuits such as the ones studied here. 

We anticipate that our strategy will enable progress in the ongoing push for the construction of quantum hardware with lower gate error rates, which will decrease the overhead cost of quantum error correction. This, in turn, will facilitate the faithful execution of more sophisticated quantum algorithms beyond the capabilities of modern classical computers, and help materialize the scientific and technological promises of the nascent second quantum revolution.         

%%%%%%%%%%%%%%%%%%%%%%%%%%%%%%%%%%%%%%%%%%
%%%%%%%%%%%%%%%%%%%%%%%%%%%%%%%%%%%%%%%%%%
%%%%%%%%%%%%%%%%%%%%%%%%%%%%%%%%%%%%%%%%%%
%%%%%%%%%%%%%%%%%%%%%%%%%%%%%%%%%%%%%%%%%%

\subsection*{Acknowledgements }
We thank M. Fishman, M. Ganahl and M. Stoudenmire for enlightening discussions. The numerical simulation were performed using the TensorFlow~\cite{tensorflow} and Qiskit~\cite{Qiskit} libraries. Numerical simulations have been performed on the Simons Foundation Super-Computing Center. This research started at the Kavli Institute for Theoretical Physics during the ``Machine Learning for Quantum Many-Body Physics'' program, and it was supported in part by the National Science Foundation under Grant No. NSF PHY-1748958. The Flatiron Institute is supported by the Simons Foundation. JC acknowledges support from Natural Sciences and Engineering Research Council of Canada (NSERC),  the Shared Hierarchical Academic Research Computing Network (SHARCNET), Compute Canada, Google Quantum Research Award, and the Canadian Institute for Advanced Research (CIFAR) AI chair program. LA acknowledges financial support from the Brazilian agencies CNPq (PQ grant No. 311416/2015-2 and INCT- IQ), FAPERJ (JCN E- 26/202.701/2018), CAPES (PROCAD2013), and the Serrapilheira Institute (grant number Serra-1709-17173) 
\bibliography{bibliography.bib}

%
%
%%%%%%%%%%%%%%------------------------------------------------%%%%%%%%%%%%%%
%%%%%%%%%%%%%%------------------------------------------------%%%%%%%%%%%%%%
%%%%%%%%%%%%%%------------------------------------------------%%%%%%%%%%%%%%
%                        			      % SUPPLEMENTARY MATERIAL %
%%%%%%%%%%%%%%------------------------------------------------%%%%%%%%%%%%%%
%%%%%%%%%%%%%%------------------------------------------------%%%%%%%%%%%%%%
%%%%%%%%%%%%%%------------------------------------------------%%%%%%%%%%%%%%
%
%
\section*{Supplementary Material}
In this supplementary material, we provide basic notions on quantum channels, a brief description of standard quantum process tomography, and details on the generation of the synthetic measurement data sets, the tensor-network representation of the Choi matrix and its reconstruction using unsupervised learning.

\section{Quantum Channels}
\begin{figure}[t!]
\noindent \centering \includegraphics[width=1\columnwidth]{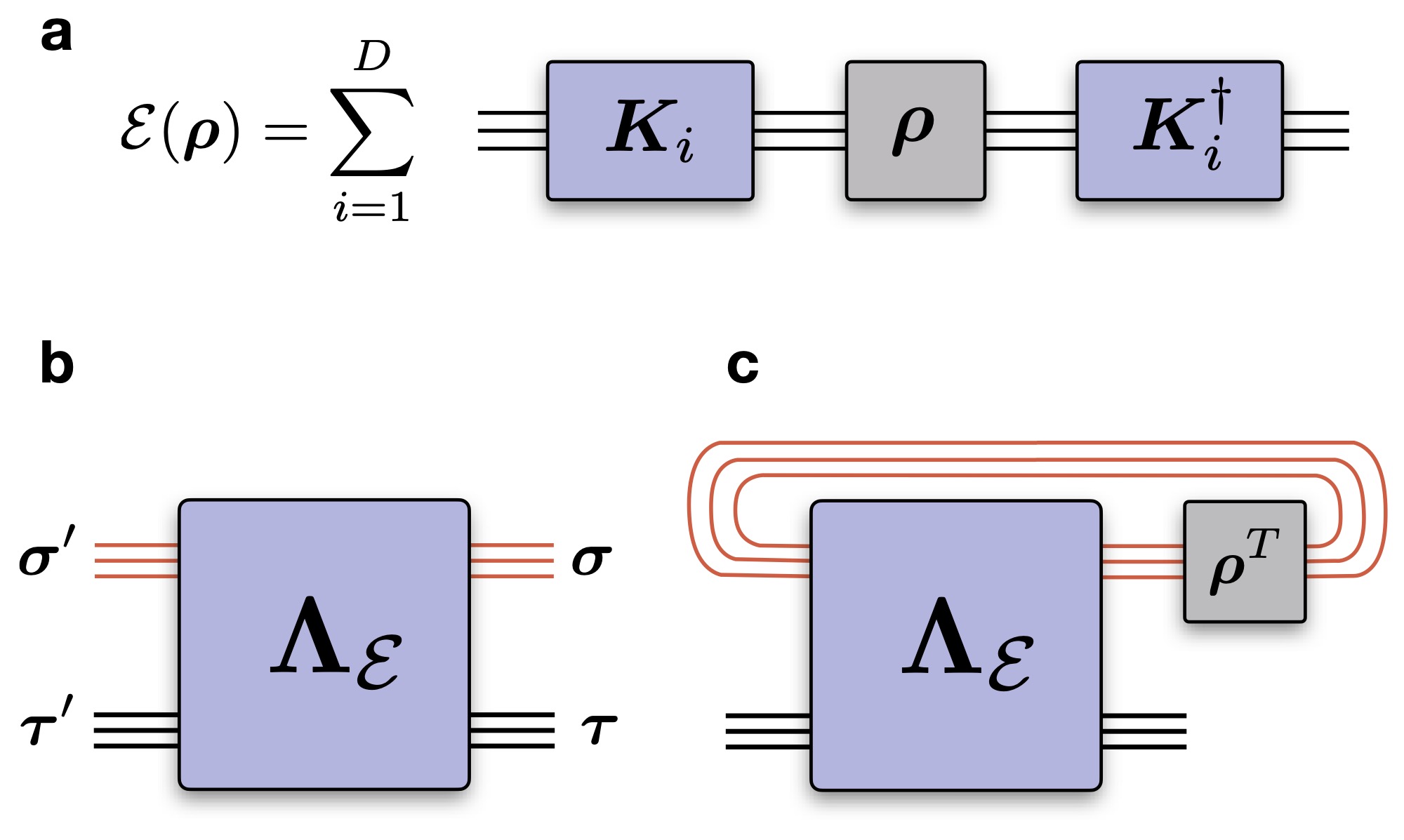}
\caption{Representations of a quantum channel (with $N=3$ qubits). ({\bf a}) Evolution of a density operator $\bm{\rho}$ under a quantum channel $\mathcal{E}$ in the Kraus representation, where the channel has decomposition over $D$ Kraus operators. ({\bf b}) Representation of the channel with the Choi matrix $\bm{\Lambda}_{\mathcal{E}}$, a rank-$4N$ tensor with the upper and lower $2N$ indices corresponding respectively to the input $\{|\bm{\sigma}\rangle\}$ and output $\{|\bm{\tau}\rangle\}$ Hilbert spaces. ({\bf c}) Evolution of $\bm{\rho}$ using Choi matrix representation. The output state of the channel $\mathcal{E}(\bm{\rho})$ is obtained by first contracting the input space $\{|\bm{\sigma}\rangle\}$ with the transpose state $\bm{\rho}^T$, followed by a trace, resulting into $\mathcal{E}(\bm{\rho})$ over the output space $\{|\bm{\tau}\rangle\}$. }
\label{Fig::SM1}
\end{figure}

The general (noisy) evolution of a quantum state with density operator $\bm{\rho}$ is described by a {\it quantum channel}, a linear map $\mathcal{E}:\;\bm{\rho}\longrightarrow\mathcal{E}(\bm{\rho})$ that is {\it completely-positive} (CP) and {\it trace-preserving} (TP). There are several equivalent representations of a CPTP map (see Ref. \cite{Wood:2015qic} for summary). One example is the {\it Kraus} representation, where the channel is expressed as a set of Kraus operators $\{\bm{K}_i\}$. The evolution of a density operator is given by (Fig.~\ref{Fig::SM1}a)
\begin{equation}
\mathcal{E}(\bm{\rho}) = \sum_{i=1}^D \bm{K}_i \bm{\rho} \bm{K}_i^\dagger\:.
\end{equation}
The CPTP property of the channel $\mathcal{E}$ constraints the Kraus operators to satisfy the completeness relation $\sum_i\bm{K}_i^\dagger\bm{K}_i=\id$.

Another representation, best suited for tomography purposes, is the {\it Choi matrix}, which for a $N$-qubit quantum channel is defined as the results of the application of the channel to the tensor product of $N$ unnormalized Bell pairs
\begin{equation}
\bm{\Lambda}_{\mathcal{E}} = (\mathbb{1}\otimes\mathcal{E}) \bigg[\bigotimes_{j=1}^N|\Phi_j^+\rangle\langle\Phi_j^+|\bigg]\:,
\end{equation}
where each Bell pair $|\Phi_j^+\rangle=\sum_{\sigma_j=\tau_j}|\sigma_j\tau_j\rangle$ is made up of an ancillary qubit $|\sigma_j\rangle$ (undergoing the channel) and a physics qubit $|\tau_j\rangle$. The channel is CP if and only if the Choi matrix is positive semi-definite, $\bm{\Lambda}_{\mathcal{E}}\ge0$. From the CP condition, it follows that the Choi matrix is isomorphic to a physical density operator over an extended Hilbert space of $2N$ qubits spanned by the basis $\{|\bm{\sigma},\bm{\tau}\rangle\}$, with ancillary qubits $|\bm{\sigma}\rangle=|\sigma_1,\dots\sigma_N\rangle$ and physical qubits $|\bm{\tau}\rangle=|\tau_1,\dots\tau_N\rangle$ (Fig.~\ref{Fig::SM1}b). The normalization of the Choi matrix is $\text{Tr}_{\bm{\sigma},\bm{\tau}}\,\bm{\Lambda}_{\mathcal{E}}=d^N$, where $d$ is the dimension of the local Hilbert space. 

The TP condition of the channel $\mathcal{E}$ requires that the partial trace of the Choi matrix over the physical qubits should yield the identity over the ancillary qubits: $\tr_{\bm{\tau}}\:\bm{\Lambda}_{\mathcal{E}}=\id_{\bm{\sigma}}$. The evolution of a generic quantum state $\bm{\rho}$ under the channel $\mathcal{E}$ is obtained through the Choi matrix as~\cite{Wood:2015qic} (Fig.~\ref{Fig::SM1}c)
\begin{equation}
\mathcal{E}(\bm{\rho})=\text{Tr}_{\bm{\sigma}}\,\Big[(\bm{\rho}^{T}\otimes\mathbb{1}_{\bm{\tau}})\bm{\Lambda}_{\mathcal{E}}\Big]\:.
\end{equation}
In this context, the ancillary and physical degrees of freedom can be interpreted respectively as input and output qubits to the channel, and will be referred to as such in the following. Finally, we note that, unlike the Kraus representation, the Choi matrix is a {\it unique} representation of the channel  $\mathcal{E}$. 

\subsection*{Tensor networks for Choi matrices}
Due to the intrinsic exponential scaling of exact classical representations of quantum states and operators, a direct estimation of the Choi matrix given the knowledge of the channel is limited to a very small number of qubits. One may wonder if, for a subset of quantum channels that admit an efficient representation (e.g. low-depth quantum circuits), the Choi matrix can be also computed efficiently. To this end, we adopt a tensor network representation of the Choi matrix, which for a $N$-qubit quantum channel is a rank-$4N$ tensor, with $2N$ input indices and $2N$ output indices~\cite{Wood:2015qic}. In its {\it canonical form}, the input and output indices are placed on the upper and lower half of the full tensor respectively (Fig.~\ref{Fig::SM1}b).

\begin{figure*}[t]
\noindent \centering \includegraphics[width=2\columnwidth]{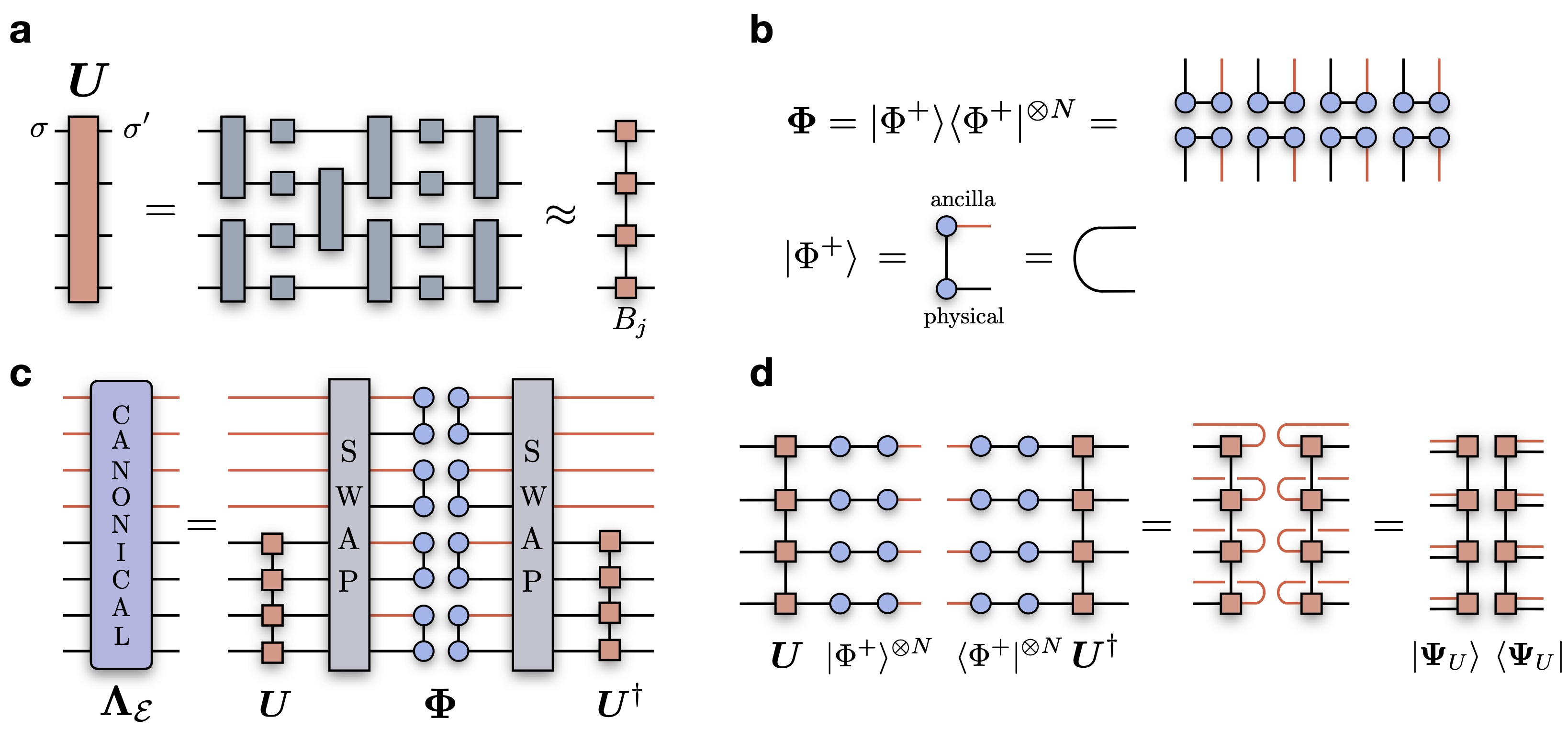}
\caption{Tensor network construction of the Choi matrix. ({\bf a}) A unitary quantum operation $\bm{U}$ (with $N=4$ qubits), compiled into a set of single- and two-qubit quantum gates. By contracting all gates together, the unitary is approximated as an MPO with bond dimension $\chi_U$. ({\bf b}) The density operator $\bm{\Phi}=|\Phi^+\rangle\langle\Phi^+|^{\otimes N}$ for the tensor product of $N$ unnormalized Bell pairs $|\Phi^+\rangle=|00\rangle+|11\rangle$. Each Bell pairs is an MPS with bond dimension $\chi_{\Phi^+}=2$, corresponding to the vectorization of the identity matrix. ({\bf c}) Canonical representation of the Choi matrix as a rank-$4N$ tensor. Within a one-dimensional tensor network representation, the canonical Choi matrix is obtained by first swapping the indices of the Bell pairs to bring all input (red) and output (black) indices together, and then contracting the resulting network with the unitary MPO on both sides. The dimension of the center bond after this operation is $2^N$. ({\bf d}) Efficient construction of the Choi matrix. Each Bell pair MPS is rearranged horizontally, corresponding to an identity matrix. The contraction with the unitary MPO is then trivial, and by folding the inner (input) indices back into each single MPO tensor, the resulting Choi matrix writes as a rank-1 density operator with physical dimension $d^2$.}
\label{Fig::SM2}
\end{figure*}

We simplify the discussion and assume that the quantum channel is noiseless, i.e. it implements a unitary evolution $\mathcal{E}:\:\bm{\rho}\longrightarrow \bm{U}\bm{\rho}\,\bm{U}^\dagger$, where $\bm{U}$ corresponds to a quantum circuits compiled into one- and two-qubit gates. Depending on the type of gates, the geometry of the circuit and its depth, the unitary $\bm{U}$ may admit an efficient representation as a matrix product operator (MPO)
\begin{equation}
\bm{U}_{\bm{\sigma}\bm{\sigma}^\prime}=\sum_{\{\bm{\mu}\}}\:
\prod_{j=1}^N\:[B_j]^{\sigma_j,\sigma_j^\prime}_{\mu_{j-1},\mu_{j}}\:,
\end{equation}
where $\bm{U}_{\bm{\sigma}\bm{\sigma}^\prime}= \langle\bm{\sigma}|\bm{U}|\bm{\sigma}^\prime\rangle$. Each $B_j$ is a rank-4 tensor with physical indices $(\sigma_j,\sigma_j^\prime)$ and {\it bond indices} $(\mu_{j-1},\mu_{j})$ (Fig.~\ref{Fig::SM2}a). The {\it bond dimension} of the MPO is defined as the maximum dimension of any bond index $\break\chi_U=\max_j\{\chi_{\mu_j}|\chi_{\mu_j}=\text{dim}[\mu_j]\}$, and it represents the measure of complexity of the MPO representation of the unitary $\bm{U}$.

The straightforward way to obtain a tensor network representation of the Choi matrix (for a unitary circuit) is to simply apply the circuit MPO to the (unnormalized) density operator $\bm{\Phi}=|\Phi^+\rangle\langle\Phi^+|^{\otimes N}$ for the $N$ unnormalized Bell pairs, each described by a matrix product state (MPS) with bond dimension $\chi_{\Phi^+}=2$ (Fig.~\ref{Fig::SM2}b). In doing so, there is freedom in how the contractions between the circuit MPO and Bell state MPS is done, stemming from different arrangements of the indices of $\bm{\Phi}$. If one were to pursue the Choi matrix in its canonical form, the MPS indices should to be properly swapped before the contraction with the MPO (Fig.~\ref{Fig::SM2}c). This however results in a very inefficient tensor network representation, as it brings the tensor product of $N$ Bell pairs into a $2N$-qubit maximally entangled state, which saturates the bond dimension of the MPS to $\chi_{\bm{\Phi}}=2^N$. 

In practice,  there is no particular reason to keep the Choi matrix in its canonical form, and an efficient representation is instead obtained as follows. First, we contract the circuit MPO with the physical indices of each Bell pair MPS (leaving out the ancilla qubits). Because the Bell state is equivalent to the vectorization of the identity matrix, the contraction between the circuit MPO and the full MPS simply returns the MPO itself. The inner (ancillary) indices can then be folded back into each local MPO tensor (Fig.~\ref{Fig::SM2}d). The result of this operation, i.e. the Choi matrix, is a rank-1 density matrix written in terms of an MPS with physical dimension $d^2$ and bond dimension $\chi_U$. Thus, the Choi matrix can be obtained efficiently as long as the MPO bond dimension is sufficiently low. Note that this operation is also called {\it unravelling} in the context of column-vectorization of dense matrices~\cite{Wood:2015qic}. In the case of a tensor product channel $\mathcal{E} = \mathcal{E}_1\otimes\dots\otimes\mathcal{E}_N$, the Choi matrix obtained in this form is the tensor product of the individual sub-system Choi matrices $\bm{\Lambda}_{\mathcal{E}}=\Lambda_{\mathcal{E}_1}\otimes\Lambda_{\mathcal{E}_2}\otimes\dots\otimes\Lambda_{\mathcal{E}_N}$. This would not be the case if the one adopted the canonical ordering of the indices (Fig.~\ref{Fig::SM1}b), leading to $\bm{\Lambda}_{\mathcal{E}}=\bm{\Lambda}_{\mathcal{E}_1\otimes\mathcal{E}_2\otimes\dots\otimes\mathcal{E}_N}$.

\begin{figure*}[t]
\noindent \centering \includegraphics[width=2\columnwidth]{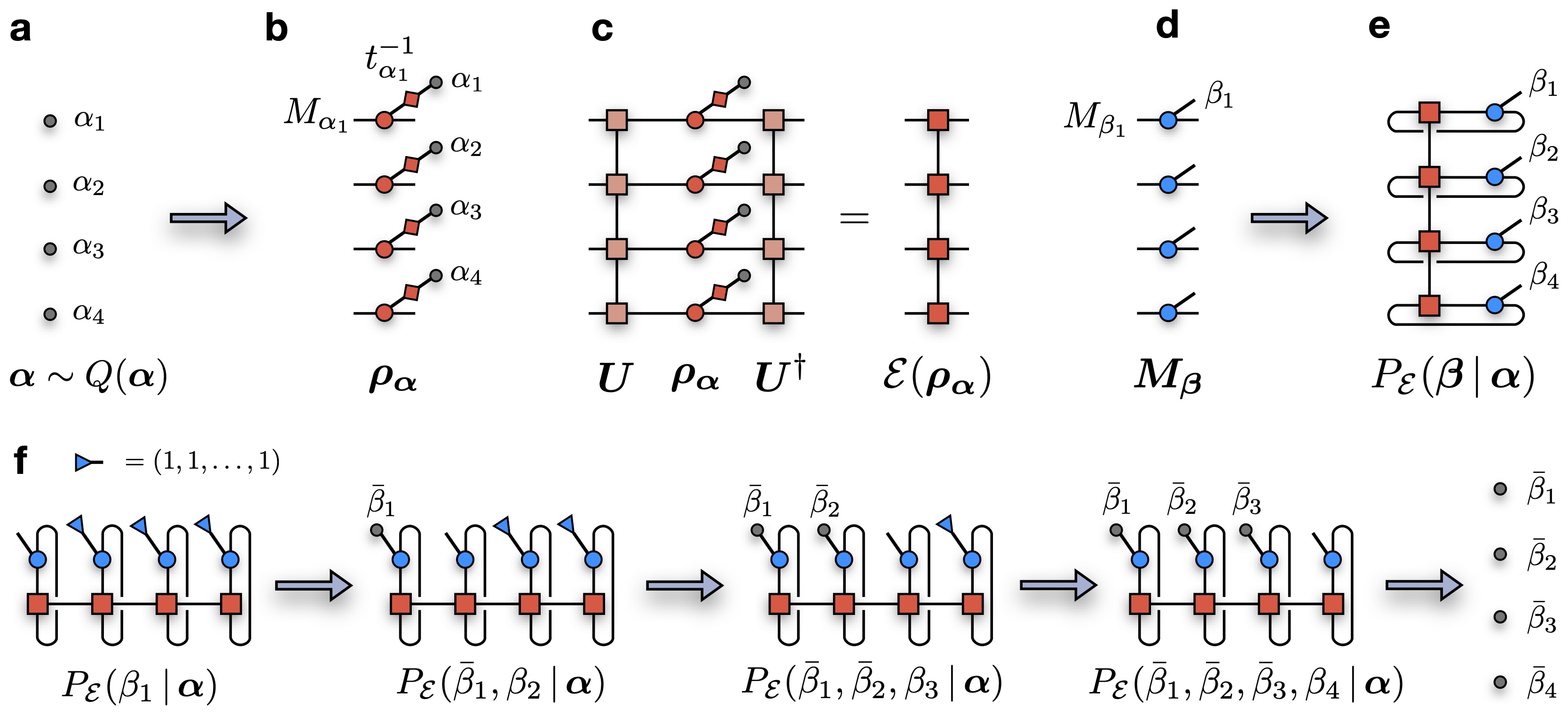}
\caption{Generation of one training data sample. ({\bf a}) First, we sample a random input POVM state $\bm{\alpha}=(\alpha_1,\alpha_2,\dots,\alpha_N)$ from a reference prior distribution $Q(\bm{\alpha})$. ({\bf b}) The string $\bm{\alpha}$ specifies an input product state $\bm{\rho}_{\bm{\alpha}}=t^{-1}_{\bm{\alpha}}\bm{M}_{\bm{\alpha}}$ to the channel. ({\bf c}) The output state of the channel is obtained by contracting the input state with the circuit MPO $\bm{U}$, resulting into a new MPO $\mathcal{E}(\bm{\rho}_{\bm{\alpha}})$. ({\bf d}) The measurement POVM $\bm{M}_{\bm{\beta}}$. ({\bf e}) The process probability distribution $P_{\mathcal{E}}(\bm{\beta}\:|\:\bm{\alpha})=\tr_{\bm{\tau}}\:[\bm{M}_{\bm{\beta}}\:\mathcal{E}(\bm{\rho_{\bm{\alpha}}})]$. ({\bf f}) Sampling scheme to obtain a single measurement outcome $\bm{\beta}$ from $P_{\mathcal{E}}(\bm{\beta}\:|\:\bm{\alpha})$. By tracing the indices $\beta_2,\dots,\beta_N$ (i.e. contracting with a vector $[1,\dots,1]$), the resulting tensor network with one open index is the probability $P_{\mathcal{E}}(\beta_1\,|\,\bm{\alpha})$, which can be sampled to generate a measurement outcome $\bar{\beta}_1$. By sweeping left to right, this procedure is repeated for each qubits, generating an outcome $\bar{\bm{\beta}}$ from the correct probability distribution $P_{\mathcal{E}}(\bm{\beta}\:|\:\bm{\alpha})$. The final result of this procedure is one single training sample $(\bm{\alpha},\bm{\beta})$. The data set are generated by repeating these steps consecutively.}
\label{Fig::SM3}
\end{figure*}

\section{Quantum process tomography}
Quantum process tomography (QPT) is a technique for reconstructing an unknown quantum channel $\mathcal{E}$ from measurement data~\cite{DAriano:2001aa}. Because of the one-to-one correspondence between the channel and its Choi matrix, QPT simply accounts of fitting the matrix elements of $\bm{\Lambda}_{\mathcal{E}}$ to the data, which consists of a special set of prepared input states to the channel and a set of measurement operators on the output states of the channel. In particular, a set of input states and measurements is called {\it informationally-complete} (IC) if the inputs $\{\bm{\rho}_{\bm{\alpha}}\}$ and the measurement operators $\{\bm{M}_{\bm{\beta}}\}$ span in full the input and the output Hilbert spaces of the quantum channel respectively. In this case, the probability distribution 
\begin{equation}
\begin{split}
P_{\mathcal{E}}(\bm{\beta}\:|\:\bm{\alpha})&=\tr_{\bm{\tau}}\:\Big[\bm{M}_{\bm{\beta}}\:\mathcal{E}(\bm{\rho_{\bm{\alpha}}})\Big]\\
&=
\tr_{\bm{\tau},\bm{\sigma}}\:\Big[(\bm{\rho}_{\bm{\alpha}}^T\otimes\bm{M}_{\bm{\beta}})\bm{\Lambda}_{\mathcal{E}}\Big]
\label{Eq::ProcessProb}
\end{split}
\end{equation}
that a measurement on the output state $\mathcal{E}(\bm{\rho}_{\bm{\alpha}})$ of the channel applied to the input state $\bm{\rho}_{\bm{\alpha}}$ yields outcome $\bm{M}_{\bm{\beta}}$ contains complete information on the channel. That is, $P_{\mathcal{E}}(\bm{\beta}\:|\:\bm{\alpha})$ uniquely characterizes the channel, and can be used to reconstruct the corresponding (unknown) Choi matrix $\bm{\Lambda}_{\mathcal{E}}$. 

The standard approach to perform QPT consists of parametrizing the Choi matrix in full (i.e. using a $4^N\times4^N$ dense matrix) and extracting its matrix elements by solving the maximum likelihood estimation problem:

\begin{align}
    \mbox{Minimize:}
        \quad& \sum_{\{(\bm{\alpha}, \bm{\beta})\}} w_{\bm{\alpha\beta}}\Big(
            P_{\mathcal{E}}(\bm{\beta}\:|\:\bm{\alpha})
            - P_{exp}(\bm{\beta}\:|\:\bm{\alpha})\Big)^2 \\
    \mbox{Subject to:}
        \quad& \bm{\Lambda}_{\mathcal{E}} \ge 0 \quad\mbox{(CP)} \\
        \quad& \tr_{\bm{\tau}}\:{\bm{\Lambda}_{\mathcal{{E}}}} = \id_{\bm{\sigma}} \quad\mbox{(TP)}.
\end{align}
where $P_{exp}(\bm{\beta}\,|\,\bm{\alpha})$ are the experimental estimates of $P_{\mathcal{E}}(\bm{\beta}\:|\:\bm{\alpha})$, and $w_{\bm{\alpha\beta}}$ are optional weights. There are two important limitations of this approach. First, it requires the parametrization of the full Choi matrix, which scales exponentially with the number of qubits. Second, in order to achieve a high-fidelity fit, the full IC set of input states and measurements is required, which also scales exponentially with $N$. For these reasons, full QPT remains restricted to very small system sizes.

\subsection*{Data sets generation}
Before discussing our algorithm for QPT, we describe how to generate the synthetic measurement data used to train the LPDOs. In our numerical experiments, we adopted, both for input states and measurement operators, the set of the rank-1 projector into the eigenstates of the Pauli matrices:
\begin{align}
    M_0 &= p_z\ketbra00, & M_1 &= p_z\ketbra11, \\
    M_2 &= p_x\ketbra++, & M_3 &= p_x\ketbra--, \\
    M_4 &= p_y\ketbra{+i}{+i}, & M_5 &= p_y\ketbra{-i}{-i}
\end{align}
In the following, we assume equal probabilities $p_x=p_y=p_z=1/3$. The full set for the $N$-qubit system is obtained from the tensor product of the operators single-qubit operators
\begin{equation}
\bm{{M}}_{\bm{\alpha}}={M}_{\alpha_1}\otimes {M}_{\alpha_2}\otimes\dots\otimes{M}_{\alpha_N}\:,
\end{equation}
and it is specified by a string $\bm{\alpha}=(\alpha_1,\dots,\alpha_N)$, with $\alpha_j=0,\dots,5$. The input states are simple product states $\bm{{\rho}}_{\bm{\alpha}}=t_{\bm{\alpha}}^{-1}\bm{{M}}_{\bm{\alpha}}$ with proper normalization $\break t_{\bm{\alpha}}=\text{Tr}\,\bm{{M}}_{\bm{\alpha}}=\prod_j\text{Tr}\,M_{\alpha_j}$. The measurement operators $\bm{M}_{\bm{\beta}}$ are defined analogously, and identified by a string $\bm{\beta}=(\beta_1,\dots,\beta_N)$.

We now provide the step-by-step procedure used to generate the training data for the case of the unitary quantum circuits. Even though the operators we implement are rank-1, we give a description for a more general case of an IC positive operator valued measures (POVM) $\bm{M}$ beyond the standard projective measurements. For a given circuit architecture, containing a set of single-qubit and two-qubit gates, we first contract each gate together to obtain the MPO corresponding to the full circuit unitary $\bm{U}$. After each application of a two-qubit gate, we restore the tensor network into an MPO structure by means of singular value decomposition. During this step, we only discard zero singular values, which implies that there is no approximation in the unitary MPO, and that the bond dimension $\chi_U$ generally grows exponentially with the depth of the circuit. 

Next, we fix a uniform prior distribution $Q(\bm{\alpha})=K^{-N}$ for the input states, where $K$ is the size of the single-qubit POVM (e.g. $K=6$ for the Pauli projectors). The POVM string $\bm{\alpha}$ is randomly sampled from $Q(\bm{\alpha})$ (Fig.~\ref{Fig::SM3}a), which defines a specific input state (Fig.~\ref{Fig::SM3}b)
\begin{equation}
\bm{\rho}_{\bm{\alpha}}=\frac{\bm{M}_{\bm{\alpha}}}{t_{\bm{\alpha}}}=
\frac{M_{\alpha_1}}{t_{\alpha_1}}\otimes\frac{M_{\alpha_2}}{t_{\alpha_2}}\otimes\dots\otimes\frac{M_{\alpha_N}}{t_{\alpha_N}}
\end{equation}
For the set of Pauli eigenstates projectors, this translates into applying one layer of single-qubit gates, according to the string $\bm{\alpha}$. The output state of the channel is then estimated by contracting $\bm{\rho}_{\bm{\alpha}}$ with the circuit MPO $\bm{U}$, $\mathcal{E}(\bm{\rho}_{\bm{\alpha}}) = \bm{U}\bm{\rho}_{\bm{\alpha}}\:\bm{U}^\dagger$ (Fig.~\ref{Fig::SM3}c). The output state $\mathcal{E}(\bm{\rho}_{\bm{\alpha}})$ is itself an MPO describing a properly normalized density operator.

Given the output state and the measurement operator $\bm{M}_{\bm{\beta}}$ (Fig.~\ref{Fig::SM3}d), the process probability $P_{\mathcal{E}}(\bm{\beta}\:|\:\bm{\alpha})$ is obtained by contracting (and tracing) these two objects together (Fig.~\ref{Fig::SM3}e). This probability can then be exactly sampled using the chain rule of probabilities~\cite{Ferris:2012aa,Carrasquilla:2019aa}. The measurement probability for qubit 1 is computed as 
\begin{equation}
p(\beta_1) = \sum_{\beta_2,\beta_3,\dots,\beta_N}p(\beta_1,\beta_2,\beta_3,\dots,\beta_N)
\end{equation}
where we introduced the short-hand notation $p(\bm{\beta})=P_{\mathcal{E}}(\bm{\beta}\:|\:\bm{\alpha})$. The probability $p(\beta_1)$ is calculated by tracing out each local POVM subspace via a contraction of the tensor network for $P_{\mathcal{E}}(\bm{\beta}\:|\:\bm{\alpha})$ with constant vectors $(1_1,1_2,\dots,1_K)$ (blue triangles) at each site $j=2,\dots,N$ (Fig.~\ref{Fig::SM3}f). Once known, the distribution can be sampled to generate measurement outcome $\bar{\beta}_1\sim P(\beta_1)$. Next, the probability distribution $p(\beta_2\:|\:\bar{\beta}_1)$ for the second qubit, conditional on the measurement of the first qubit, is calculated as the ratio between $p(\bar{\beta}_1,\beta_2)$ (shown in the second network of Fig.~\ref{Fig::SM3}f) and $p(\bar{\beta}_1)$. By repeating this procedure, one obtains a final configuration $\bm{\bar{\beta}}$ sampled from the correct probability distribution $p(\bm{\beta}) = P_{\mathcal{E}}(\bm{\beta}\:|\:\bm{\alpha})$. Importantly, each $N$-qubit measurement outcome is completely uncorrelated from any other.

For the noisy quantum channels studied in the paper, since there are only $N=5$ qubits, we perform a direct simulation of the channel to obtain the full Choi matrix. The training data is obtained directly from the Choi matrix, using input states and measurement operators identical to the ones described above.

\begin{figure}[t!]
\noindent \centering \includegraphics[width=1\columnwidth]{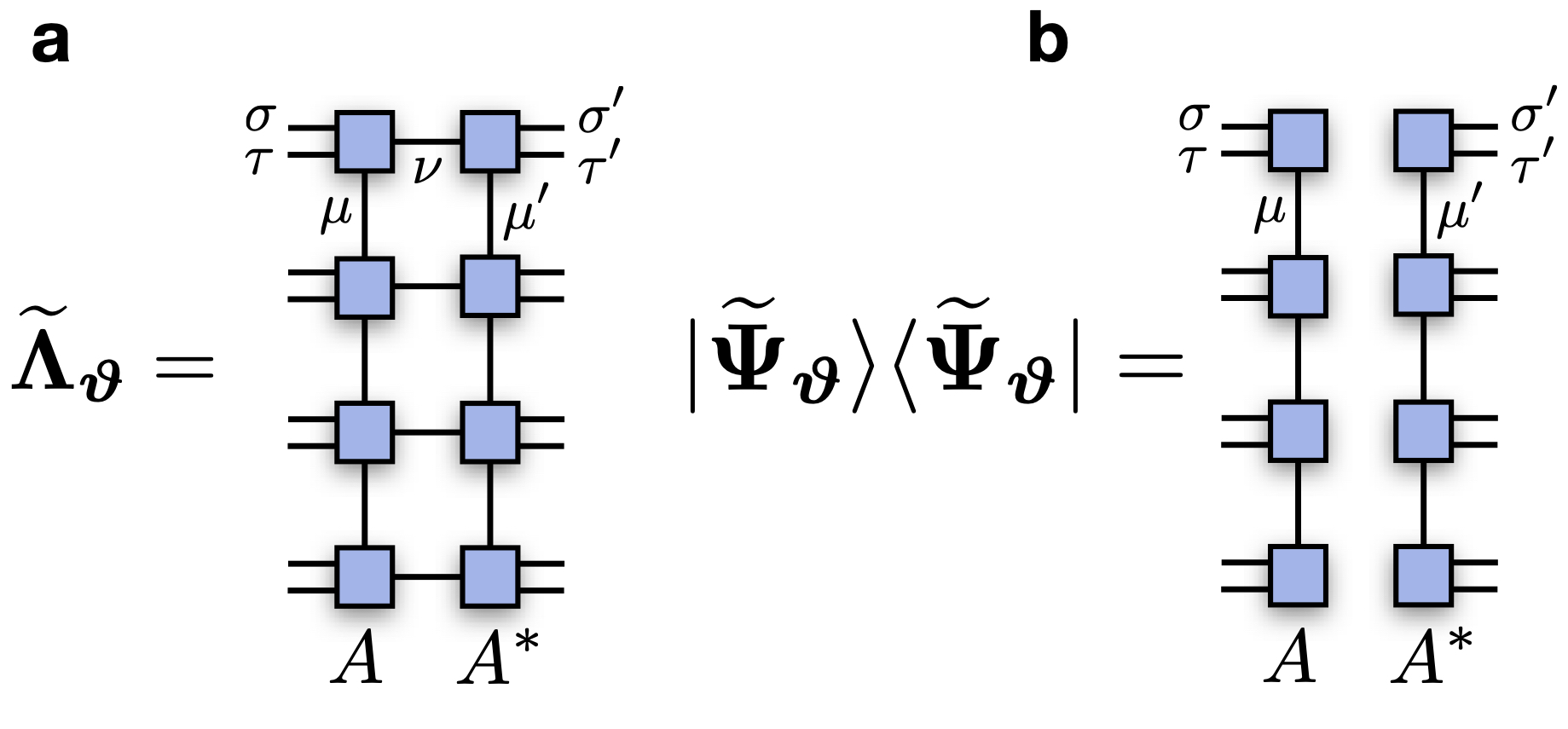}
\caption{Locally-purified density operators. ({\bf a}) A LPDO $\widetilde{\bm{\Lambda}}_{\bm{\vartheta}}$ representing a $N=4$ qubit process, with 4 input indices $\{\sigma_j\}$, 4 output indices $\{\tau_j\}$. We group input and output indices together, leading to a representation of the Choi matrix in terms of 4 LPDO rank-4 tensors $\{\widetilde{A}_j\}$ (rank-3 for boundary tensors). ({\bf b}) A rank-1 LPDO with Kraus dimension $\chi_\nu=1$ $\widetilde{\bm{\Lambda}}_{\bm{\vartheta}}=|\widetilde{\Psi}_{\bm{\vartheta}}\rangle\!\langle\widetilde{\Psi}_{\bm{\vartheta}}|$, where $\widetilde{\Psi}_{\bm{\vartheta}}$ is an MPS.}
\label{Fig::SM4}
\end{figure}

\subsection*{Locally-purified density operators}
We start by defining the parametrization of the Choi matrix in terms of a {\it locally-purified density operator} (LPDO)~\cite{Werner:2016aa}. In the input/output basis defined before, the matrix elements of the unnormalized LPDO are
\begin{equation}
[\widetilde{\bm{\Lambda}}_{\bm{\vartheta}}]^{\bm{\tau},\bm{\tau}^\prime}_{\bm{\sigma},\bm{\sigma}^\prime}=\sum_{\{\bm{\mu},\bm{\mu}^\prime\}}\sum_{\{\bm{\nu}\}}\:
\prod_{j=1}^N\:[\widetilde{A}_j]^{\tau_j,\sigma_j}_{\mu_{j-1},\nu_j,\mu_{j}}[\widetilde{A}^*_j]^{\tau^\prime_{j},\sigma^\prime_j}_{\mu^\prime_{j-1},\nu_j,\mu^\prime_{j}}\:,
\label{Eq::LPDO}
\end{equation}
where each tensor $\widetilde{A}_j$ has input index $\sigma_j$, output index $\tau_j$, bond indices $(\mu_j,\mu_{j+1})$ and {\it Kraus index} $\nu_j$ (Fig.~\ref{Fig::SM4}a). The bond and Kraus dimensions of the LPDO are defined as $\chi_\mu=\max_j\{\chi_{\mu_j}=\text{dim}[\mu_j]\}$ and $\chi_\nu=\max_j\{\chi_{\nu_j}=\text{dim}[\nu_j]\}$.

By construction, the Choi matrix $\widetilde{\bm{\Lambda}}_{\bm{\vartheta}}$ is positive, $\widetilde{\bm{\Lambda}}_{\bm{\vartheta}}\ge0$ and Hermitian $\widetilde{\bm{\Lambda}}_{\bm{\vartheta}}=\widetilde{\bm{\Lambda}}^\dagger_{\bm{\vartheta}}$. The normalization $Z_{\bm{\vartheta}}=\text{Tr}_{\bm{\sigma},\bm{\tau}}\:\widetilde{\bm{\Lambda}}_{\bm{\vartheta}}$ can be computed with cost $O(Nd^2\chi_\nu\chi_\mu^3)$. If the Kraus dimension is set to $\chi_\nu=1$, the Choi matrix reduces to rank-1, and writes $\widetilde{\bm{\Lambda}}_{\bm{\vartheta}}=|\widetilde{\bm{\Psi}}_{\bm{\vartheta}}\rangle\!\langle\widetilde{\bm{\Psi}}_{\bm{\vartheta}}|$ Fig.~\ref{Fig::SM4}b), where 
$\widetilde{\bm{\Psi}}_{\bm{\vartheta}}$ is a matrix product state (MPS) 
\begin{equation}
[\widetilde{\bm{\Psi}}]_{\bm{\sigma}}^{\bm{\tau}}=\sum_{\{\bm{\mu}\}}\:
\prod_{j=1}^N\:[\widetilde{A}_j]^{\tau_j,\sigma_j}_{\mu_{j-1}\mu_{j}}\:.
\label{Eq::MPS}
\end{equation}
with physical dimension $d^2$ and bond dimension $\chi_\mu$.

\begin{figure*}[t]
\noindent \centering \includegraphics[width=2\columnwidth]{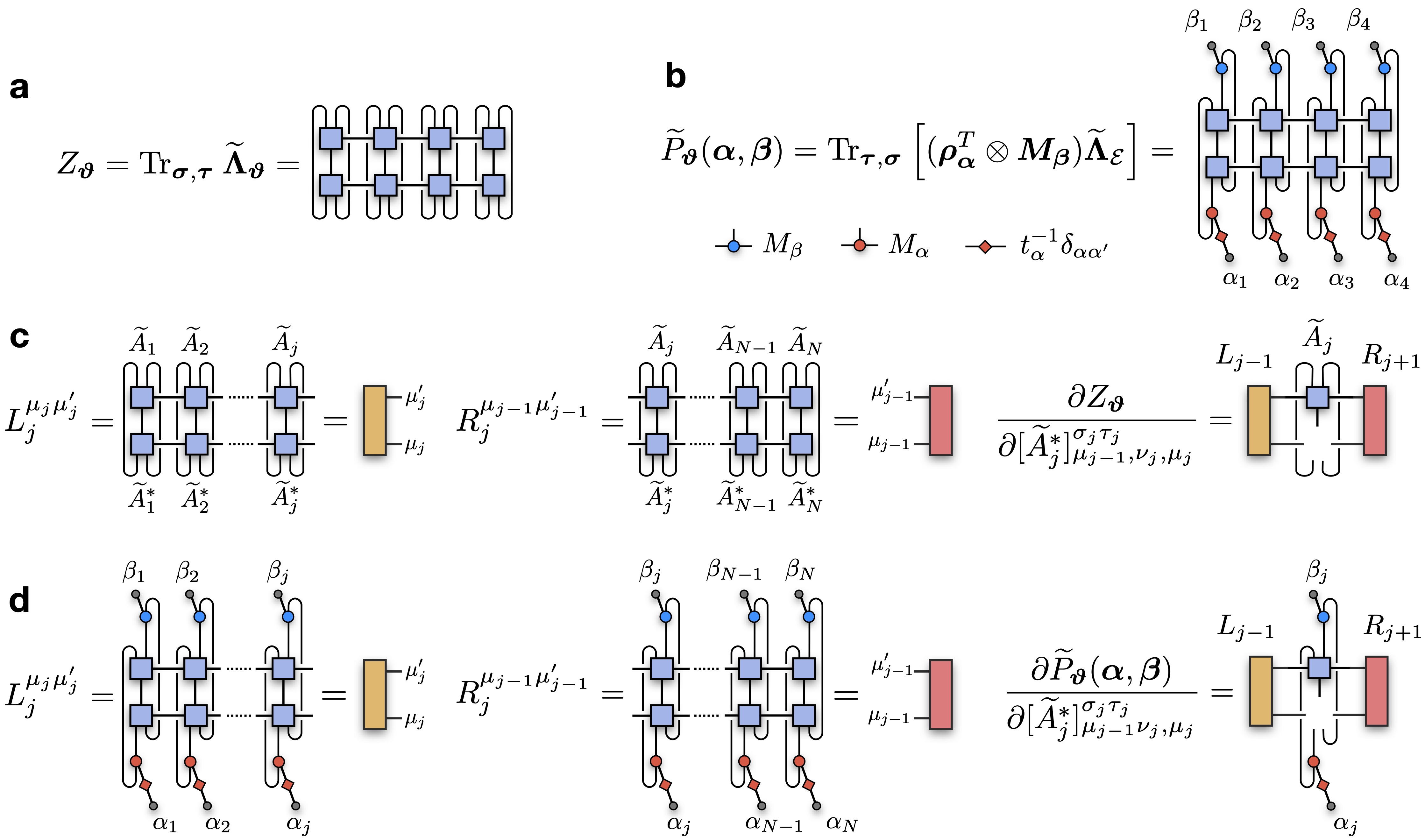}
\caption{Tensor-network optimization. ({\bf a}) Tensor network contraction to evaluate the normalization $Z_{\bm{\vartheta}}$ of the LPDO. ({\bf b}) Tensor network contraction to evaluate the unnormalized probability $\widetilde{P}_{\bm{\vartheta}}(\bm{\beta}\:|\:\bm{\alpha})$. ({\bf c}) Calculation of the gradients of the normalization $\partial_{\bm{\vartheta}}Z_{\bm{\vartheta}}$. First, the left-environment tensors $L_j$ ($j=1,\dots,N-1$) are computed sequentially (and stored) by contracting the Kraus index and tracing the input/output index of the LPDO, from left to right. The same is repeated by contracting from right to left for the right-environment tensors $R_j$ ($j=2,\dots,N$). Then, the gradients of the normalization with respect to the conjugate tensor $\widetilde{A}^*_j$ are evaluated using with previously stored environment tensors. ({\bf d}) Calculation of the gradients of the unnormalized probability $\partial_{\bm{\vartheta}}\widetilde{P}_{\bm{\vartheta}}(\bm{\alpha},\bm{\beta})$ for one given data point $(\bm{\alpha},\bm{\beta})$. As for the normalization, left and right environment tensors are calculated -- where now the LPDO is contracted with input and output states corresponding to $\bm{\alpha}$ and $\bm{\beta}$ respectively -- and subsequently used to evaluate the gradients with respect to each conjugate tensor $\widetilde{A}^*_j$. }
\label{Fig::SM5}
\end{figure*}

\subsection*{Unsupervised learning}
The parameters of the LPDO -- the tensor components $\bm{\vartheta}=\{\widetilde{A}_j\}$ -- are variationally optimized by minimizing the Kullbach-Leibler (KL) divergence
\begin{equation}
\mathcal{D}_{KL}=\sum_{\{\bm{\alpha}\}}Q(\bm{\alpha})\sum_{\{\bm{\beta}\}} P_{\mathcal{E}}(\bm{\beta}\,|\,\bm{\alpha})\log\frac{P_{\mathcal{E}}(\bm{\beta}\,|\,\bm{\alpha})}{P_{\bm{\vartheta}}(\bm{\beta}\,|\,\bm{\alpha})}\:.
\label{Eq::KL_full_SM}
\end{equation}
where $P_{\mathcal{E}}(\bm{\beta}\,|\,\bm{\alpha})$ is the process probability defined in Eq.~(\ref{Eq::ProcessProb}). The probability distribution $P_{\bm{\vartheta}}(\bm{\beta}\,|\,\bm{\alpha})$ associated to the process described by the LPDO is
\begin{equation}
\begin{split}
P_{\bm{\vartheta}}(\bm{\beta}\:|\:\bm{\alpha})&=
\tr_{\bm{\tau},\bm{\sigma}}\:\Big[(\bm{\rho}_{\bm{\alpha}}^T\otimes\bm{M}_{\bm{\beta}})\bm{\Lambda}_{\bm{\vartheta}}\Big]\\
&=Z_{\bm{\vartheta}}^{-1}d^N\tr_{\bm{\tau},\bm{\sigma}}\:\Big[(\bm{\rho}_{\bm{\alpha}}^T\otimes\bm{M}_{\bm{\beta}})\widetilde{\bm{\Lambda}}_{\bm{\vartheta}}\Big]\\
&\equiv Z_{\bm{\vartheta}}^{-1}\widetilde{P}_{\bm{\vartheta}}(\bm{\beta}\:|\:\bm{\alpha})\:,
\end{split}
\end{equation}
where $\bm{\Lambda}_{\bm{\vartheta}}=d^NZ_{\bm{\vartheta}}^{-1}\widetilde{\bm{\Lambda}}_{\bm{\vartheta}}$ is the properly normalized LPDO Choi matrix, and we defined the unnormalized LPDO probability distribution $\widetilde{P}_{\bm{\vartheta}}(\bm{\beta}\:|\:\bm{\alpha})$. By averaging Eq.~(\ref{Eq::KL_full_SM}) over the data set $\mathcal{D}$, we obtain the negative log-likelihood
\begin{equation}
\begin{split}
\mathcal{C}({\bm{\vartheta}})&=-\frac{1}{M}\sum_{k=1}^{M}\log P_{\bm{\vartheta}}(\bm{\beta}_k\,|\,\bm{\alpha}_k)\\
&=\log Z_{\bm{\vartheta}} - \frac{1}{M}\sum_{k=1}^{M}\log \widetilde{P}_{\bm{\vartheta}}(\bm{\beta}_k\,|\,\bm{\alpha}_k)\\
&= \log Z_{\bm{\vartheta}} -\big\langle \log\widetilde{P}_{\bm{\vartheta}}(\bm{\beta}\,|\,\bm{\alpha}) \big\rangle_{\mathcal{D}}\:,
\label{Eq::NLLSM}
\end{split}
\end{equation}
where we omitted the constant entropy term of the target distribution.

Given the cost function $\mathcal{C}({\bm{\vartheta}})$, the LPDO parameters are tuned according to the gradients
\begin{equation}
\begin{split}
\mathcal{G}_{\bm{\vartheta}} &= \frac{\mathcal{C}({\bm{\vartheta}})}{\partial\bm{\vartheta}}=\frac{\partial}{\partial\bm{\vartheta}} \log Z_{\bm{\vartheta}} - \frac{\partial}{\partial\bm{\vartheta}} \big\langle \log\widetilde{P}_{\bm{\vartheta}}(\bm{\beta}\,|\,\bm{\alpha}) \big\rangle_{\mathcal{D}}\\
&=\frac{1}{Z_{\bm{\vartheta}}} \frac{\partial Z_{\bm{\vartheta}}}{\partial\bm{\vartheta}} 
-\bigg\langle\frac{1}{\widetilde{P}_{\bm{\vartheta}}(\bm{\beta}\,|\,\bm{\alpha})}
\frac{\partial \widetilde{P}_{\bm{\vartheta}}(\bm{\beta}\,|\,\bm{\alpha})}{\partial \bm{\vartheta}}\bigg\rangle_{\mathcal{D}}\:.
\end{split}
\end{equation}
Since, in general, the tensor components $\{\widetilde{A}_j\}$ are complex-valued, one should adopt the Wirtinger derivatives, and update each tensor $\widetilde{A}_j$ with the gradient taken with respect to its conjugate value $\widetilde{A}^*_j$.

The calculation of the gradients proceeds in two steps~\cite{Han:2018aa,glasser2019expressive}. First one evaluates the normalization $Z_{\bm{\vartheta}}$ with a trace over all the input/output indices of the LPDO (Fig.~\ref{Fig::SM5}a). The gradient of $Z_{\bm{\vartheta}}$ with respect to the component $\widetilde{A}_j^*$ corresponds to the tensor network used to compute $Z_{\bm{\vartheta}}$ with the tensor $\widetilde{A}_j^*$ removed from it. To reduce the number of tensor contractions required to compute the full set of gradients, one should first calculate and store the set of {\it environment tensors} $\{L_j\}$ and $\{R_j\}$ shown in Fig.~\ref{Fig::SM5}c, obtained in two sweeps over the LPDO respectively from left to right and from right to left. The gradients of the normalization with respect to each tensor $\widetilde{A}_j^*$ are then calculated with a third sweep as 
\begin{equation}
\frac{\partial Z_{\bm{\vartheta}}}{\partial[\widetilde{A}^*_j]^{\sigma_j\tau_j}_{\mu_{j-1},\nu_j,\mu_{j}}}=
\sum_{\mu_j^\prime,\mu^\prime_{j-1}} L_{j-1}^{\mu_{j-1},\mu^\prime_{j-1}}[\widetilde{A}_j]^{\sigma_j\tau_j}_{\mu^\prime_{j-1},\nu_j,\mu^\prime_{j}}R_{j+1}^{\mu_j,\mu^\prime_j}\:.
\end{equation}

The second step repeats this procedure for the data-dependent term in the cost function. For each single data point $(\bm{\alpha},\bm{\beta})$, one computes the unnormalized probability $\widetilde{P}_{\bm{\vartheta}}(\bm{\beta}\:|\:\bm{\alpha})$ by contracting the LPDO with the corrrersponding input state $\bm{\rho}_{\bm{\alpha}}$ and measurement operator $\bm{M}_{\bm{\beta}}$ (Fig.~\ref{Fig::SM5}b). One then sweeps through the LPDO to accumulate the left and right environment tensors, and compute the gradient of the unnormalized probability analogously to the normalization (Fig.~\ref{Fig::SM5}d). The final gradients are simply the average over all data samples.

\subsection*{Trace-preserving regularization}

\begin{figure}[t!]
\noindent \centering \includegraphics[width=1\columnwidth]{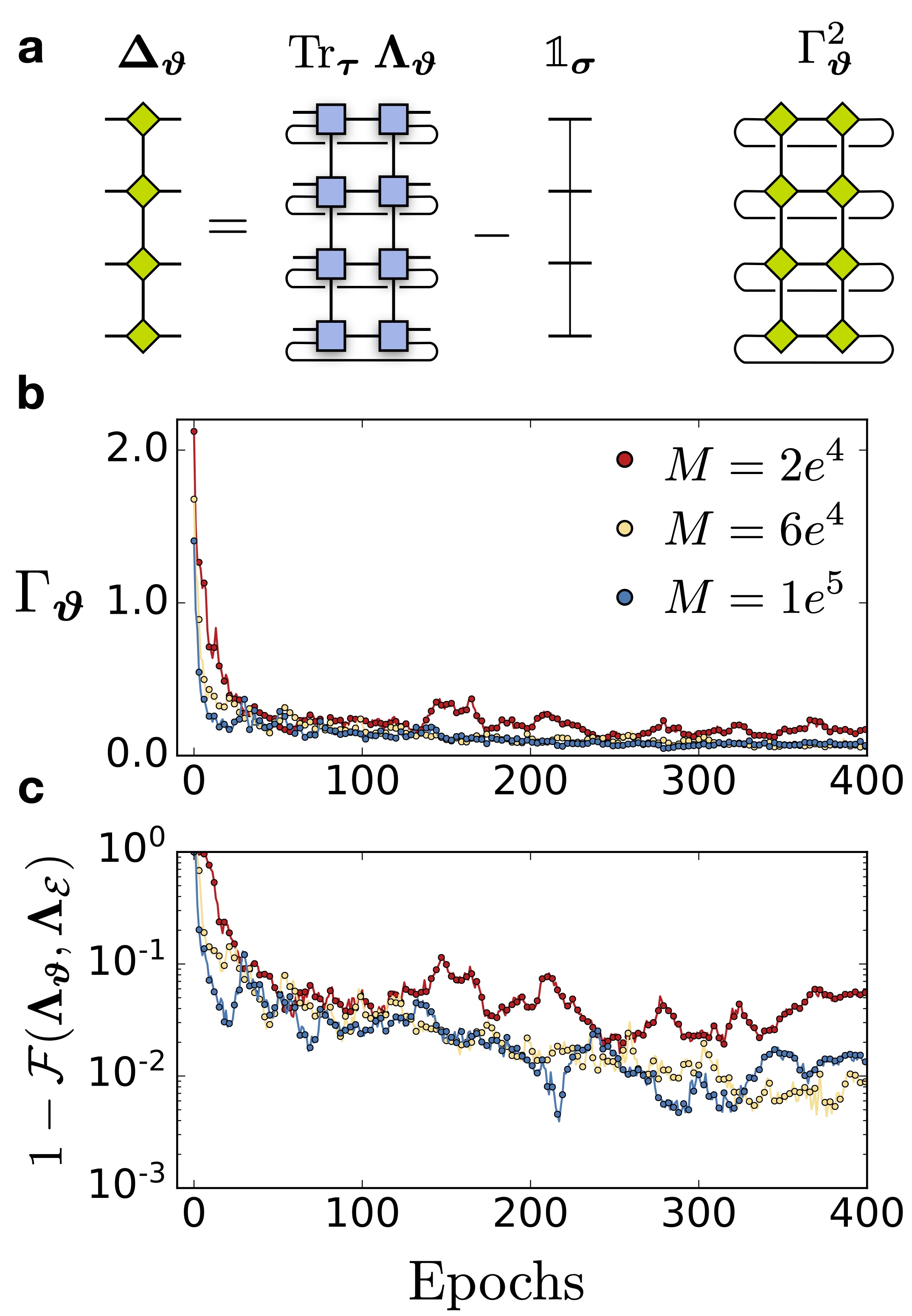}
\caption{Trace-preserving regularization. ({\bf a}) Tensor network for $\bm{\Delta}_{\bm{\vartheta}}$, obtained by subtracting the identity MPO (with bond dimension 1) to the (properly normalized) LPDO $\bm{\Lambda}_{\bm{\vartheta}}$, and tensor contraction required to compute $\Gamma_{\bm{\vartheta}}$. We show the measurement of the regularization $\Gamma_{\bm{\vartheta}}$ ({\bf b}) and the reconstruction infidelity ({\bf c}) during the training for a one-dimension random quantum circuit with $N=10$ qubits and depth $D=2$, for different number of the total data set size $M$. }
\label{Fig::SM6}
\end{figure}

Finally, we note that the LPDO, in general, does not enforce the TP condition on the corresponding quantum channel, i.e. $\text{Tr}_{\bm{\tau}}\,\bm{\Lambda}_{\bm{\vartheta}}\ne\mathbb{1}_{\bm{\sigma}}$. This condition can be easily added to the cost function as a {\it regularization term}, which biases the optimization to yield a set of optimal parameters $\bm{\vartheta}$ that minimizes the negative log-likelihood, while also minimizing the {\it distance} between $\text{Tr}_{\bm{\tau}}\,\bm{\Lambda}_{\bm{\vartheta}}$ and $\mathbb{1}_{\bm{\sigma}}$. As a distance measure, we choose the Frobenius norm of the difference $\bm{\Delta}_{\bm{\vartheta}}=\text{Tr}_{\bm{\tau}}\bm{\Lambda}_{\bm{\vartheta}}-\mathbb{1}_{\bm{\sigma}}$:
\begin{equation}
\|\bm{\Delta}_{\bm{\vartheta}}\|_F=\sqrt{\text{Tr}_{\bm{\sigma}}\big(\bm{\Delta}_{\bm{\vartheta}}\bm{\Delta}_{\bm{\vartheta}}^\dagger\big)}\:.
\end{equation}
The tensor network for $\bm{\Delta}_{\bm{\vartheta}}$ can be easily computed by performing an MPO subtraction~\cite{Hubig17}, which in this case it increases the bond dimension of $\bm{\Lambda}_{\bm{\vartheta}}$ by 1 (Fig.~\ref{Fig::SM6}a). The regularization term is then
\begin{equation}
\Gamma_{\bm{\vartheta}}=\sqrt{d^{-N}}\sqrt{\text{Tr}_{\bm{\sigma}}\big(\bm{\Delta}_{\bm{\vartheta}}\bm{\Delta}_{\bm{\vartheta}}^\dagger\big)}\:,
\end{equation}
where we introduced a normalization pre-factor $\sqrt{d^{-N}}$. This leads to the final cost function
\begin{equation}
\mathcal{C}(\bm{\vartheta})=\log Z_{\bm{\vartheta}} -\big\langle \log\widetilde{P}_{\bm{\vartheta}}(\bm{\beta}\,|\,\bm{\alpha}) \big\rangle_{\mathcal{D}} + \kappa \Gamma_{\bm{\vartheta}}\:,
\end{equation}
where $\kappa$ is an additional hyper-parameter. 

We show the measurement of the regularization term $\Gamma_{\bm{\vartheta}}$ (Fig.~\ref{Fig::SM6}b) at each training iteration for the reconstruction of one-dimensional random quantum circuits of different depths. By comparing these curves with the reconstruction infidelities (Fig.~\ref{Fig::SM6}c), one can clearly see the correlation between the accuracy of the reconstruction and the amount of violation of the TP condition.

\subsection*{Gradient updates}
Once the gradients $\mathcal{G}_{\bm{\vartheta}}$ with respect to each tensor components are known, the LPDO is updated using gradient descent. In its simplest form, the parameters are changed according to 
\begin{equation}
\bm{\vartheta}\longrightarrow\bm{\vartheta}-\eta\:\mathcal{G}_{\bm{\vartheta}}\:,
\end{equation}
where $\eta$ is the size of the update (i.e. the {\it learning rate}), and the gradients are computed over the full training data set:
\begin{equation}
\mathcal{G}_{\bm{\vartheta}} = -\frac{1}{M}\sum_{k=1}^M\frac{\partial}{\partial\bm{\vartheta}}\log P_{\bm{\vartheta}}(\bm{\beta}_k\:|\:\bm{\alpha}_k)\:.
\end{equation}
This type of gradient update results however into a very slow training, as the number of training samples $M$ might be large. In practice, the gradients are computed on a {\it batch} of data containing $M_B\ll M$ training samples. One training {\it epoch} is then defined as a sweep of the full data set $\mathcal{D}$ (being reshuffled at its start), with the parameters $\bm{\vartheta}$ being updated $M/M_B$ times. The advantage, aside a faster training, is that the fluctuations induced in the gradients due to the smaller number of samples $M_B$ helps the optimization to escape local minima.

The other hyper-parameter of the optimization is the learning rate $\eta$. A major limitation of the vanilla gradient descent update shown above is that choosing the correct value of $\eta$ in advance can be difficult. Further, the learning rate is identical for all parameters. This problem is resolved by using more advanced optimizations schemes. We specificallu use the Adam optimizer (from {\it Adaptive Moment Estimation})~\cite{kingma2014adam}, where each parameter $\vartheta_k$ is updated with an adaptive learning rate $\eta_k$. For the set of parameters $\bm{\vartheta}^{(t)}$ at a given epoch $t$, the Adam optimizer estimates moving averages of first and second moments of the gradients
\begin{align}
\mathcal{M}^{(t)}_{1,k} &= \xi_1\mathcal{M}^{(t-1)}_{1,k} + (1-\xi_1)\;\mathcal{G}_{\vartheta_k}\\
\mathcal{M}^{(t)}_{2,k} &= \xi_2\mathcal{M}^{(t-1)}_{2,k} + (1-\xi_2)\;\mathcal{G}^2_{\vartheta_k}\:,
\end{align}
where $\xi_1$ and $\xi_2$ are hyper-parameters controlling the rate of decay of the moving averages. Each parameters is then updated as
\begin{equation}
\vartheta_k^{(t+1)}=\vartheta_k^{(t)}-\eta\frac{\widehat{\mathcal{M}}^{(t)}_{1,k}}{\sqrt{\widehat{\mathcal{M}}^{(t)}_{2,k}}+\epsilon}\:,
\end{equation}
where $\epsilon$ is used to avoid numerical instabilities, and $\widehat{\mathcal{M}}^{(t)}_{i,k}=\mathcal{M}^{(t)}_{i,k}/(1-\xi_i)$ are used to correct the bias introduced by the zero-initialization $\mathcal{M}^{(t=0)}_{i,k}=0$~\cite{kingma2014adam}.

\subsection*{Overfitting and model selection}
The goal of training the LPDO using unsupervised learning is to efficiently extract the relevant structure and features characterizing the unknown channel from a limited set of measurements. In other words, the model needs to be able to {\it generalize} beyond the measurements provided for its training. If the number of samples in the data set $\mathcal{D}$ is too low, it is likely that the LPDO training leads to {\it overfitting}, i.e. the LPDO learns features present in the data that are not representative of the unknown channel, but only stems from the limited number of training samples.

A strategy to monitor the overfitting, routinely used in the training of deep neural networks, is to divide the data set into two sub-sets: a training data set $\mathcal{D}_T$ and a validation data set $\mathcal{D}_V$. Here, we do so using a 80\%/20\% split ratio. The training data set $\mathcal{D}_T$ is used for the learning procedure, i.e. the calculation of the gradients used to update the model. During training, we compute the training loss (i.e. the average of the cost function on the training data set)
\begin{equation}
\mathcal{L}_T(\bm{\vartheta})=-\frac{1}{|\mathcal{D}_T|}\sum_{(\bm{\alpha},\bm{\beta})\in\mathcal{D}_T}\log P_{\bm{\vartheta}}(\bm{\beta}\:|\:\bm{\alpha})\:,
\end{equation}
which signals whether the model is actively learning (i.e. a decreasing $\mathcal{L}_T(\bm{\vartheta})$). At the same time, we also compute the validation loss on the held-out data
\begin{equation}
\mathcal{L}_V(\bm{\vartheta})=-\frac{1}{|\mathcal{D}_V|}\sum_{(\bm{\alpha},\bm{\beta})\in\mathcal{D}_V}\log P_{\bm{\vartheta}}(\bm{\beta}\:|\:\bm{\alpha})\:.
\end{equation}
Here, $|\mathcal{D}_T|$ and $|\mathcal{D}_V|$ are the size of the training and validation data sets respectively. 

Generally, in the early stage of the training, the validation loss decreases hand-in-hand with the training loss. However, if the model starts to overfit spurious features in the training data, the validation loss will invert its trend and start increasing, an indication that more training data is needed. We stress that both of these measurements are available in a practical experimental setting, since no information about the channel is being used.

The validation loss $\mathcal{L}_V(\bm{\vartheta})$ is also a useful metric to perform the model selection, i.e. to pick a specific set of parameters $\bm{\vartheta}^{(t)}$ at epoch $t$ to be considered the optimal solution of the optimization problem. In our numerical simulations, we select the optimal parameters as the ones at the training epochs $t$ where the measurement of the validation loss returned its lowest value. This is also a model selection procedure that can be used in an experimental setting.

\subsection*{Fidelity estimation}
Once the unknown quantum channel $\mathcal{E}$ with Choi matrix $\bm{\Lambda}_{\mathcal{E}}$ has been reconstructed, the learned Choi matrix $\bm{\Lambda}_{\bm{\vartheta}}$ (for a set of parameters $\bm{\vartheta}$ selected as shown above) can be used for certification. The metric we adopted is the quantum process fidelity, defined as the quantum state fidelity between Choi matrices 
\begin{equation}
\mathcal{F}(\bm{\Lambda}_{\bm{\vartheta}},\bm{\Lambda}_{\mathcal{E}})=d^{-2N}
\bigg(\text{Tr}\sqrt{\sqrt{\bm{\Lambda}_{\mathcal{E}}}\bm{\Lambda}_{\bm{\vartheta}}\sqrt{\bm{\Lambda}_{\mathcal{E}}}}\bigg)^2\:.
\end{equation} 
In general, this measurement is not scalable and remains restricted to small system sizes. However, it is also not a useful metric in a practical setting, where the Choi matrix $\bm{\Lambda}_{\mathcal{E}}$ is not known. In practice, the reconstruction fidelity is computed using the target Choi matrix $\bm{\Lambda}_{\mathcal{E}}=|\bm{\Psi}_{\mathcal{E}}\rangle\!\langle\bm{\Psi}_{\mathcal{E}}|$ for the ideal unitary implemented by the quantum circuit
\begin{equation}
\mathcal{F}(\bm{\Lambda}_{\bm{\vartheta}},\bm{\Lambda}_{\mathcal{E}})=d^{-2N}
\langle\bm{\Psi}_{\mathcal{E}}|\bm{\Lambda}_{\bm{\vartheta}}|\bm{\Psi}_{\mathcal{E}}\rangle\:.
\end{equation}
For the specific case where the LPDO Choi matrix is rank-1, this reduces to
\begin{equation}
\mathcal{F}(\bm{\Lambda}_{\bm{\vartheta}},\bm{\Lambda}_{\mathcal{E}})=d^{-2N}|\langle\bm{\Psi}_{\mathcal{E}}|\bm{\Psi}_{\bm{\vartheta}}\rangle|^2\:.
\end{equation}

\subsection*{Specifics of the numerical experiments}
In this final section, we provide details on the numerical experiments presented in the main text. In all cases, the LPDO tensors $\{\widetilde{A}_j\}$ are initialized randomly, with each tensor component set to 
\begin{equation}
[\widetilde{A}_j]^{\tau_j,\sigma_j}_{\mu_{j-1},\nu_j,\mu_{j}} = a_r + i a_i
\end{equation}
where $a_r$ and $a_i$ are drawn from a uniform distribution centered around zero with width 0.2. We compute the gradients on batches of data containing $M_B=800$ samples. Once the gradients are collected, we update the LPDO tensors using the Adam optimization with parameters $\eta=0.005$, $\xi_1=0.9$, $\xi_2=0.999$ and $\epsilon=10^{-7}$.

\paragraph{Figure 2.}
The first set of quantum channels investigated are unitary quantum circuits containing one layer of single-qubit gates. We study two types of circuits, containing either Hadamard gates
\begin{equation}
H=\frac{1}{\sqrt{2}}\begin{pmatrix}1 & 1\\1 & -1\end{pmatrix}\:,
\end{equation}
or random single-qubit rotations
\begin{equation}
R(\theta,\phi,\lambda)=\begin{pmatrix}
\cos \frac{\theta}{2} & -e^{i\lambda}\sin\frac{\theta}{2}\\
e^{i\phi}\sin\frac{\theta}{2} & e^{i(\phi+\lambda)}\cos\frac{\theta}{2}
\end{pmatrix}\:.
\end{equation}
To obtain the sample complexity curves shows in Fig.~\ref{Fig::2}b, we perform the reconstruction for an increasing number $N$ of qubits. For each $N$, we start using a small data set size $M$, and increase it with a fixed size-step until the threshold $\varepsilon=0.025$ in infidelity is met. The result is a value $M^*$ with an error bar given by the size-step. 

We repeat the same scaling study for quantum circuits containing $D$ layers of controlled-NOT (CX) gates
\begin{equation}
\text{CX} = \begin{pmatrix}
1 & 0 & 0 & 0 \\
0 & 1 & 0 & 0 \\
0 & 0 & 0 & 1 \\
0 & 0 & 1 & 0 \\
\end{pmatrix}\:.
\end{equation}
For a quantum circuit with depth $D$, the odd and even layers apply two-qubit gates with the control qubit having odd and even qubit-index respectively. Here, the bond dimension of the LPDO Choi matrix is set to the bond dimension of the circuit MPO.

\paragraph{Figure 3.}
We reconstruct random quantum circuits in both one and two dimensions. In both cases, each layer of the quantum circuit consist of one layer with $N$ single-qubit random rotations $R(\theta,\phi,\lambda)$ (defined above) and one layer of CX gates. In the one-dimensional geometry, the CX gates alternates as in the previous case. For the two-dimensional quantum circuit, they are applied according to the color scheme shown in Fig.~\ref{Fig::3}b. For the simulation of the quantum circuit and the data generation, the circuit MPO has a ``snake-shape'' as per usual in MPS simulations of two-dimensional geometries. After applying the CX gates, the circuit tensor network is restored into a local form by means of singular value decomposition, where only zero singular values are discarded. This means that the representation of the target quantum circuit is exact.

We first set of the bond dimension of the LPDO Choi matrix equal to the bond dimension of the circuit MPO, and set the Kraus dimension to $\chi_\nu=1$. All the data shown in Fig.~\ref{Fig::3} has been collected under this condition. However, additional simulations have also been performed using larger values of the LPDO bond dimension, obtaining comparable results. During the training, we monitor the training loss, the validation loss, the TP regularizer, and the reconstruction fidelity. We use cross-validation on the held-out data set $\mathcal{D}_V$ to select the best models for each circuit configuration and for each data set size $M$. The curves in Fig.~\ref{Fig::3}e-f show the reconstruction infidelities of these selected models.

\paragraph{Figure 4.}
Finally, we reconstruct a noisy quantum channel. We consider the X-stabilizer measurement in the surface code, where the parity-check between four data qubits is measured using an additional (measurement) qubit with the quantum circuit shown in Fig.~\ref{Fig::4}a. The circuit contains two Hadamard gates and four CX gates. We apply an amplitude damping channel, characterized by the Kraus operators
\begin{align}
K_0&=|0\rangle\langle0| + \sqrt{1-\gamma}\:|1\rangle\langle1|\\
K_1&=\sqrt{\gamma}\:|0\rangle\langle1|\:
\end{align}
where $\gamma$ is the decay probability. The channel is applied to each quantum gate in the circuit, where for the two-qubit gates the channel is just the tensor product of the single-qubit channel shown above.

We now relax any prior information on both the quantum circuit and the noise channel. We perform the reconstruction by varying the bond dimension $\chi_\mu$ and the Kraus dimension $\chi_\nu$ of the LPDO. The only setting where convergence in the training metrics is found already for $\chi_\nu=0$ is the noiseless channel $\gamma=0$, as expected. Nonetheless, even by increasing $\chi_\nu$, the noiseless channel is still properly reconstructed. This can be seen in Fig.~\ref{Fig::4}b, where the purity of the reconstruction LPDO Choi matrix for $\gamma=0$ and $\chi_\nu=6$ reaches the correct value of $\text{Tr}\bm{\Lambda}_{\bm{\vartheta}}\approx1$. The infidelity curves are shown for a fixed data set size of $M=5\times10^5$.

\end{document}